\documentclass[conference]{IEEEtran}

\ifCLASSINFOpdf
\else
\fi
\usepackage[export]{adjustbox}
\usepackage{graphicx}
\usepackage{tabularx,booktabs}
\usepackage{dingbat}
\usepackage{diagbox}
\usepackage{multirow} 
\usepackage[hyphens]{url}
\usepackage[hidelinks,breaklinks]{hyperref}
\usepackage{xcolor}
\usepackage{amsfonts, amsmath, amsthm, amssymb}
\usepackage{subcaption}
\hypersetup{breaklinks=true}
\usepackage{tikz}
\usepackage{textcomp}
\usepackage{lipsum}
\usepackage{setspace}
\usepackage{listings}
\usepackage{color}
\usepackage{svg}
\usepackage[noabbrev,capitalize]{cleveref}

\definecolor{dkgreen}{rgb}{0,0.6,0}
\definecolor{gray}{rgb}{0.5,0.5,0.5}
\definecolor{mauve}{rgb}{0.58,0,0.82}

\IEEEoverridecommandlockouts
\newcommand\copyrighttext{%
  \footnotesize \textcopyright 2024 IEEE.  Personal use of this material is permitted.  Permission from IEEE must be obtained for all other uses, in any current or future media, including reprinting/republishing this material for advertising or promotional purposes, creating new collective works, for resale or redistribution to servers or lists, or reuse of any copyrighted component of this work in other works.}
\newcommand\copyrightnotice{%
\begin{tikzpicture}[remember picture,overlay]
\node[anchor=south,yshift=10pt] at (current page.south) {\fbox{\parbox{\dimexpr\textwidth-\fboxsep-\fboxrule\relax}{\copyrighttext}}};
\end{tikzpicture}%
}

\usepackage{graphicx}
\graphicspath{{figures/}}
\begin{document}

%
\title{Evaluation of Hardware-based Video Encoders on Modern GPUs for UHD Live-Streaming}

\author{
    \IEEEauthorblockN{Kasidis Arunruangsirilert, Jiro Katto}
    \IEEEauthorblockA{Department of Computer Science and Communications Engineering, Waseda University, Tokyo, Japan}
    \IEEEauthorblockA{\{kasidis, katto\}@katto.comm.waseda.ac.jp}
}
%
\maketitle
\copyrightnotice
\begin{abstract}

Many GPUs have incorporated hardware-accelerated video encoders, which allow video encoding tasks to be offloaded from the main CPU and provide higher power efficiency. Over the years, many new video codecs such as H.265/HEVC, VP9, and AV1 were added to the latest GPU boards. Recently, the rise of live video content such as VTuber, game live-streaming, and live event broadcasts, drives the demand for high-efficiency hardware encoders in the GPUs to tackle these real-time video encoding tasks, especially at higher resolutions such as 4K/8K UHD. In this paper, RD performance, encoding speed, as well as power consumption of hardware encoders in several generations of NVIDIA, Intel GPUs as well as Qualcomm Snapdragon Mobile SoCs were evaluated and compared to the software counterparts, including the latest H.266/VVC codec, using several metrics including PSNR, SSIM, and machine-learning based VMAF. The results show that modern GPU hardware encoders can match the RD performance of software encoders in real-time encoding scenarios, and while encoding speed increased in newer hardware, there is mostly negligible RD performance improvement between hardware generations. Finally, the bitrate required for each hardware encoder to match YouTube transcoding quality was also calculated.

\end{abstract}




%
\IEEEpeerreviewmaketitle

\vspace{-1mm}
\section{Introduction}
 \vspace{-0.5mm}
 \setstretch{0.97}
\looseness=-1
With the rise of multimedia content, modern Graphics Processing Unit (GPU) as well as Mobile System on a Chip (SoC) manufacturers such as Intel, NVIDIA, and Qualcomm have allocated part of the semiconductor die to implement a hardware core designed specifically to tackles video decoding and encoding task since 2011 \cite{7477508}. While AV1, one of the latest video codecs, was introduced over six years ago, the hardware implementation of such codec takes several years to catch up. Only recently did the hardware manufacturers start to introduce hardware-accelerated AV1 encoding in their product, with Intel first introducing AV1 encoding in their ARC GPU line-up in mid-2022, in late-2022 for NVIDIA as part of their Ada Lovelace GPU, and finally, Qualcomm just recently introduced such capability in their Snapdragon 8 Gen 3 Mobile SoCs during the late-2023. Due to the nature of application-specific integrated circuits (ASIC), these hardware encoders in modern GPU can deliver very high encoding speed at relatively low power consumption, but due to the space and complexity limitations when designing a semiconductor, some of the encoder's functionalities such as support of Bi-directional predicted frame (B Frames), the maximum number of reference frames, and Rate-Distortion Optimization (RDO), were either omitted or modified in some hardware implementation \cite{6737724}. This resulted in differences in rate-distortion (RD) performance between each hardware implementation, which usually yields worse RD performance when compared to software implementations such as libx264 (AVC), libx265 (HEVC), libvpx-vp9 (VP9), and SVT-AV1 (AV1).

Recently, the evolution of network infrastructure and Radio Access Networks (RAN) \cite{10118777} resulted in the transmission of High-Definition (HD) and Ultra-High-Definition (UHD) live video content via the internet to become mainstream. While modern high-end desktop CPUs are powerful enough to perform real-time HD video encoding to mainstream codecs such as H.264/AVC via software encoder with reasonably high-quality presets, the lower-end and mobile CPUs might not be powerful enough to handle such tasks. Furthermore, many online content creators such as Game Streamers and Virtual YouTubers (VTubers) usually have a single PC or a single laptop setup, which requires the video game and in the case of VTuber, an additional Live2D Model, to be run aside the encoding tasks. By using the CPU for the video encoding, there will be a major performance impact on the system as the CPU is getting overwhelmed by the encoding task, leaving little to no CPU headroom for other tasks \cite{7965585}. Additionally, the power consumption of the software encoder is usually excessive as the CPU is required to operate in a high-power state \cite{10109035}, which may not be suitable for live broadcasting from battery-powered devices such as laptop computers and smartphones. Finally, due to the increased complexity of modern video codecs, which can deliver a superior RD performance over H.264/AVC, such as H.265/HEVC, VP9, and AV1 \cite{8456303}, in many cases even modern high-end CPUs couldn't keep up with real-time encoding to these codecs \cite{9240439}, especially at 4K/8K UHD resolution with high frame rate of 60 or 120 fps. While hardware encoders couldn't match the encoding efficiency of the best quality presets of software encoders, modern hardware-accelerated video encoders implementation in GPUs can achieve a similar RD performance to medium-quality presets, which are usually employed in real-time encoding, of software encoders. Therefore, there is a high demand for high-efficiency hardware-accelerated video encoders to handle the real-time encoding of live video content for transmission over online streaming platforms as these encoders are the key to realizing real-time content creation at UHD resolutions.

In this paper, the RD performance (encoding efficiency), encoding speed, and power consumption of various hardware-accelerated encoders from various manufacturers will be evaluated and compared to software counterparts, at various encoding bitrates, and codecs including the most common H.264/AVC, H.265/HEVC, VP9, and recently practically implemented AV1. The RD performance difference between each generation of hardware from each manufacturer will also be investigated. Since the main focus of this paper is real-time encoding for online distribution, the encoding parameters will follow the recommended encoding parameters from popular online live-streaming services including YouTube, Twitch, and Twitcasting, to ensure that the results in this work reflect the actual real-world result in this use case scenario. Furthermore, the encoding quality of YouTube will also be evaluated, to find out an optimized encoding bitrate for each codec at each resolution. Finally, various types of content such as VTuber Promotional Video (PV), Japanese animation, film, and documentary, both in Standard Dynamic Range (SDR) and High Dynamic Range (HDR) at resolutions of 1080p and 4K will be encoded using Intel Arc A770 GPU, which delivers the best RD performance among GPU encoders \cite{MSU_2023}, then compared to our datasets to give the reader a more throughout context on how each encoder will perform on their genre of content. Finally, the bitrate required to match the quality of YouTube transcodes is provided to help gamers, streamers, and VTubers decide an appropriate bitrate for their live encoders. This paper is organized as follows. Section II discusses the experiment setup including the hardware, dataset, as well as quality measurement procedure. Next, the experiment results are presented and analyzed in Section III. Then, a conclusion is provided in Section IV. Finally, the future work is discussed in Section V. Our main contributions are listed below: \looseness=-1
\begin{itemize}
\item We demonstrated that software encoders couldn't provide sufficient encoding throughput at UHD resolution even on a high-end consumer desktop CPU, despite consuming more power than the hardware-accelerated counterpart.
\item We evaluated and compared several different hardware-accelerated video encoders implemented in modern GPU using two datasets: Twitch and ITE.
\item To provide the reader with context on our dataset, we provide the RD curve of our dataset compared to real-world contents at two different resolutions.
\item Unlike the existing works, the main focus of this study is real-time encoding. Therefore, the encoding parameters used were adjusted and optimized for real-time encoding, which resulted in RD performance impact.
\item RD performance between hardware generations from each manufacturer were also found and evaluated.
\end{itemize}

\section{Experiment Setup}
\vspace{-1mm}

\subsection{Encoders}

For NVIDIA hardware, six hardware with different micro-architectures with NVENC support were used including Maxwell 1st Generation, Maxwell 2nd Generation, Pascal, Turing, Ampere, and Ada Lovelace. As for Intel, three integrated graphics and one dedicated graphics processor with Intel QuickSync support were used. Finally, for Qualcomm mobile SoC cases, three 5G smartphones with three different generations of Qualcomm Snapdragon SoCs were used. The hardware utilized in this work can be summarized in Table \ref{tab:hardware}. To access the hardware encoder FFmpeg was used. NVIDIA hardware was accessed using the NVENC encoder, the QSV encoder was used for Intel hardware, and Qualcomm hardware was accessed using the MediaCodec encoder provided by Android OS. As for software encoders, the encoders used for each codec in this paper can be summarized in Table \ref{tab:software}. For the system resource consumption experiment, which has a target resolution and real-time frame rate target of 1080p60, a Dell Precision 5680 Laptop with Intel Core i7-13700H processor was used. The laptop was set to Ultra Performance Mode with the CPU Power Limit removed, which allows the CPU to draw sustained power of 70W. On the other hand, the encoding speed with 2160p as the target resolution was evaluated on a desktop computer with an AMD Ryzen 9 5900X processor. It should be noted that the rest of the RD performance experiments were done on several different hardware to speed up the process. Except for Android-based encoder cases, all encodings were done on Microsoft Windows 10 or 11 operating systems. The power consumption experiment was carried out on a laptop as total system power consumption can be obtained directly from the system agent due to the whole system being integrated onto one board, whereas on a desktop PC, the power consumption must be measured via a wattmeter. The Full HD resolution was chosen for the laptop experiment as the CPU couldn't keep up with the real-time encoding of UHD video. However, for the encoding speed experiment, a desktop PC with AMD Ryzen 9 5900X CPU was used as it's more practical to use a software encoder on a desktop processor due to its extremely high computational power requirement, which resulted in sustained high power consumption and heat output. This usually results in thermal throttling in many laptops due to the limitation in heat dissipation capacity, causing the result to vary wildly from one run to another. Hardware decoders on the same die as the encoders were used and NVIDIA hardware decoders were used for software encoding.

\begin{table}[!tbp]
\setstretch{0.85}
\caption{Software Encoders used for evaluation}
\vspace{-1.5mm}
\centering
\label{tab:software}
\resizebox{7.5cm}{!}{\begin{tabular}{@{}lll@{}}
\toprule
Codecs               & Encoder  & Version\\\midrule
FFmpeg & & N-111929-gbfa43447fa-20230906\\
H.264/AVC & libx264 & core 164\\
H.265/HEVC & libx265 & 3.5+103-8f18e3ad3\\
H.266/VVC & Fraunhofer VVenC & v1.9.1\\
VP9 & libvpx-vp9 & v1.13.0 \\
AV1 & SVT-AV1 & v1.6.0-4-g903ff3ad\\
\bottomrule
\end{tabular}}
\vspace{-7mm}
\end{table}

\vspace{-1mm}

\subsection{Encoding Parameters}

\begin{table*}[!tbp]
\setstretch{0.83}
\caption{VMAF score when Spatial AQ and Temporal AQ are enabled and disabled on NVIDIA NVENC Encoders}
\vspace{-2mm}
\centering
\label{tab:TemporalAQCompare}
\resizebox{18cm}{!}{\begin{tabular}{@{}lc ccc ccc ccc ccc ccc ccc@{}}
\toprule

\multirow{4}{*}{Dataset}& \multirow{4}{*}{Resolution} & \multicolumn{6}{c}{H.264/AVC} & \multicolumn{6}{c}{H.265/HEVC} & \multicolumn{6}{c}{AV1} \\
\cmidrule(lr){3-8}\cmidrule(lr){9-14}\cmidrule(l){15-20}
& & \multicolumn{3}{c}{Enabled} & \multicolumn{3}{c}{Disabled} & \multicolumn{3}{c}{Enabled} & \multicolumn{3}{c}{Disabled}& \multicolumn{3}{c}{Enabled} & \multicolumn{3}{c}{Disabled} \\
\cmidrule(lr){3-5}\cmidrule(lr){6-8}\cmidrule(lr){9-11}\cmidrule(lr){12-14}\cmidrule(lr){15-17}\cmidrule(l){18-20}
&&Low&Medium&High&Low&Medium&High&Low&Medium&High&Low&Medium&High&Low&Medium&High&Low&Medium&High\\
\midrule
ITE 4K & 1080p/FHD & 60.91&72.36&79.36&\textbf{\textcolor{blue}{64.72}}&\textbf{\textcolor{blue}{74.70}}&\textbf{\textcolor{blue}{80.79}}&68.26&75.76&80.34&\textbf{\textcolor{blue}{69.29}}&\textbf{\textcolor{blue}{76.49}}&\textbf{\textcolor{blue}{80.89}}&69.23&76.99&81.74&\textbf{\textcolor{blue}{69.94}}&\textbf{\textcolor{blue}{77.46}}&\textbf{\textcolor{blue}{82.05}}\\
& 2160p/4K & 67.39&76.74&82.45&\textbf{\textcolor{blue}{71.10}}&\textbf{\textcolor{blue}{79.19}}&\textbf{\textcolor{blue}{84.13}}&72.35&78.80&82.73&\textbf{\textcolor{blue}{73.70}}&\textbf{\textcolor{blue}{79.90}}&\textbf{\textcolor{blue}{83.68}}&73.90&80.15&83.98&\textbf{\textcolor{blue}{74.95}}&\textbf{\textcolor{blue}{81.03}}&\textbf{\textcolor{blue}{84.74}} \\
\midrule
Twitch & 1080p/FHD & 65.86&79.61&88.01&\textbf{\textcolor{blue}{70.59}}&\textbf{\textcolor{blue}{81.89}}&\textbf{\textcolor{blue}{88.79}}&76.10&84.32&89.33&\textbf{\textcolor{blue}{76.86}}&\textbf{\textcolor{blue}{84.88}}&\textbf{\textcolor{blue}{89.78}}&77.82&85.68&90.48&\textbf{\textcolor{blue}{78.19}}&\textbf{\textcolor{blue}{85.92}}&\textbf{\textcolor{blue}{90.65}}\\
& 2160p/4K & 75.05&86.18&92.97&\textbf{\textcolor{blue}{78.85}}&\textbf{\textcolor{blue}{88.14}}&\textbf{\textcolor{blue}{93.81}}&82.78&89.53&93.65&\textbf{\textcolor{blue}{83.83}}&\textbf{\textcolor{blue}{90.34}}&\textbf{\textcolor{blue}{94.32}}&84.54&90.82&94.65&\textbf{\textcolor{blue}{85.16}}&\textbf{\textcolor{blue}{91.30}}&\textbf{\textcolor{blue}{95.05}} \\
\bottomrule
\end{tabular}}
\vspace{-5mm}
\end{table*}

Online live-streaming platforms usually publish the required parameters for live encoders. To ensure that the results in this paper are the best representation of real-world performance and are compatible with popular streaming services, the encoding guidelines of YouTube \cite{google} and Twitch \cite{twitch} were followed. All of the video encoders were configured to encode the video in constant bitrate mode (CBR) with a keyframe interval of two seconds. While CBR doesn't give the best RD performance, it ensures that there is no spike in TCP throughput due to changes in scene complexity, which prevents issues like throughput starvation, frame dropping, and buffering, especially on consumer-grade internet connections. For these reasons, Twitch strictly requires that all ingest streams must be encoded using the CBR mode, or it will be rejected. For H.264/AVC encoding, the high profile was used. Some encoders that require a target constant rate factor (CRF) were configured to target a CRF value of 18. The buffer size was set to double the value of the target bitrate. For the supported encoder, the maximum number of B-Frame was set to two, and the maximum number of reference frames was set to one. For NVIDIA NVENC, the highest quality preset (P7) was used with multipass encoding enabled and a lookahead of 120 frames. Spatial Adaptive Quantization (AQ) and Temporal Adaptive Quantization (AQ) are both disabled as they may negatively impact the objective quality metrics \cite{nvidia_2023} as seen in \cref{tab:TemporalAQCompare}. For Intel hardware, the very slow preset was used with a look ahead of 100 frames. Nine different target bitrates, which vary depending on the resolution, were used and can be summarized in Table \ref{tab:bitrate}. The commands used to encode the video can also be found below:

\begin{lstlisting}
// NVIDIA NVENC
// H.264/AVC
ffmpeg -c:v h264_cuvid -i input.mp4 -vcodec h264_nvenc -profile:v high -refs 1 -preset 18 -rc cbr -tune 1 -spatial-aq 0 -temporal-aq 0 -multipass 2 -rc-lookahead 120 -bf 2 -b_adapt 1  -b:v {Target Bitrate} -g 120 output.mp4
// H.265/HEVC
ffmpeg -c:v h264_cuvid -i input.mp4 -vcodec hevc_nvenc -refs 1 -preset 18 -rc cbr -tune 1 -spatial-aq 0 -temporal-aq 0 -multipass 2 -rc-lookahead 120 -bf 2 -b_adapt 1 -an -b:v {Target Bitrate} -g 120 output.mp4
// AV1
ffmpeg -c:v h264_cuvid -i input.mp4 -vcodec av1_nvenc -refs 1 -preset 18 -rc cbr -tune 1 -spatial-aq 0 -temporal-aq 0 -multipass 2 -rc-lookahead 120 -bf 2 -b_adapt 1 -an -b:v {Target Bitrate} -g 120 output.mp4

// Intel QuickSync Video
// H.264/AVC
ffmpeg -c:v h264_qsv -i input.mp4 -vcodec h264_qsv -profile:v high -refs 1 -bf 2 -rc cbr -preset 1 -an -minrate {Target Bitrate} -maxrate {Target Bitrate} -b:v {Target Bitrate} -bufsize {2*Target Bitrate} -g 120 output.mp4
// H.265/HEVC
ffmpeg -c:v h264_qsv -i input.mp4 -vcodec hevc_qsv -refs 1 -bf 2 -rc cbr -preset 1 -an -minrate {Target Bitrate} -maxrate {Target Bitrate} -b:v {Target Bitrate} -bufsize {2*Target Bitrate} -g 120 output.mp4
// VP9
ffmpeg -c:v h264_qsv -i input.mp4 -vcodec vp9_qsv -refs 1 -bf 2 -rc cbr -preset 1 -an -minrate {Target Bitrate} -maxrate {Target Bitrate} -b:v {Target Bitrate} -bufsize {2*Target Bitrate} -g 120 output.mp4
// AV1
ffmpeg -c:v h264_qsv -i input.mp4 -vcodec av1_qsv -refs 1 -bf 2 -rc cbr -preset 1 -an -minrate {Target Bitrate} -maxrate {Target Bitrate} -b:v {Target Bitrate} -bufsize {2*Target Bitrate} -g 120 output.mp4

// Android MediaCodec
// H.264/AVC
ffmpeg -c:v h264_mediacodec -i input.mp4 -vcodec h264_mediacodec -bitrate_mode cbr -an -b:v {Target Bitrate} -g 120 output.mp4
// H.265/HEVC
ffmpeg -c:v h264_mediacodec -i input.mp4 -vcodec hevc_mediacodec -bitrate_mode cbr -an -b:v {Target Bitrate} -g 120 output.mp4

//Software Encoder
// libx264
ffmpeg -i input.mp4 -vcodec libx264 -profile:v high -rc cbr -preset {Preset} -crf 18 -rc-lookahead 120 -bf 2 -x264-params "b_pyramid=2:bframes=2:ref=1" -minrate {Target Bitrate} -maxrate {Target Bitrate} -b:v {Target Bitrate} -bufsize {2*Target Bitrate} -g 120 output.mp4
// libx265
ffmpeg -i input.mp4 -vcodec libx265 -profile:v main -rc cbr -preset {Preset} -crf 18 -rc-lookahead 120 -bf 2 -x265-params "b_pyramid=2:bframes=2:ref=1" -minrate {Target Bitrate} -maxrate {Target Bitrate} -b:v {Target Bitrate} -bufsize {2*Target Bitrate} -g 120 output.mp4
// libvpx-vp9
ffmpeg -i input.mp4 -vcodec libvpx-vp9 -rc cbr -deadline realtime -cpu-used {Preset} -crf 18 -rc-lookahead 120 -bf 2 -refs 1 -minrate {Target Bitrate} -maxrate {Target Bitrate} -b:v {Target Bitrate} -bufsize {2*Target Bitrate} -g 120 output.mp4
// libsvtav1
ffmpeg -i input.mp4 -vcodec libsvtav1 -tier main -rc cbr -preset {Preset} -crf 18 -rc-lookahead 120 -bf 2 -refs 1 -an -minrate {Target Bitrate} -maxrate {Target Bitrate} -b:v {Target Bitrate} -bufsize {2*Target Bitrate} -g 120 output.mp4
// libvvenc
ffmpeg -i input.mp4 -vcodec libvvenc -rc-lookahead 120 -bf 2 -refs 1 -b:v {Target Bitrate} -g 120 -preset {Preset} -vvenc-params threads=32 -g 120 output.mp4

\end{lstlisting}
\vspace{-1mm}
\begin{table}[!tbp]
\setstretch{0.85}
\caption{Hardware used for the evaluations}
\vspace{-1.5mm}
\centering
\label{tab:hardware}
\resizebox{8.5cm}{!}{\begin{tabular}{@{}ll@{}}
\toprule
Test Case                 & Hardware  \\\midrule
NVIDIA Maxwell 1st Generation & NVIDIA Quadro M1200 \\
NVIDIA Maxwell 2nd Generation & NVIDIA GeForce GTX TITAN X \\
NVIDIA Pascal & NVIDIA GeForce GTX 1080 \\
NVIDIA Turing & NVIDIA GeForce RTX 2080 SUPER \\
NVIDIA Ampere & NVIDIA GeForce RTX 3090 \\
NVIDIA Ada Lovelace & NVIDIA RTX 2000 Ada Generation Laptop\\
Intel 6-9th Generation Processor & Intel HD Graphics 630 via Intel Core i7-7820HQ \\
Intel 10-11th Generation Processor & Intel UHD Graphics 750 via Intel Core i7-11700 \\
Intel 12-13th Generation Processor & Intel Iris Xe Graphics via Intel Core i7-13700H \\
Intel Arc Graphics & Intel Arc A770 16 GB \\
Qualcomm Snapdragon 888 & ASUS Smartphone for Snapdragon Insiders\\
Qualcomm Snapdragon 8 Gen 1 & Samsung Galaxy S22 Ultra 5G (SC-52C)\\
Qualcomm Snapdragon 8 Gen 2 & Samsung Galaxy Z Flip5 (SM-F731B)\\
\bottomrule
\end{tabular}}
\vspace{-2mm}
\end{table}

\begin{table}[!tbp]
\setstretch{0.85}
\caption{Encoding bitrate used for each resolution}
\vspace{-1.5mm}
\centering
\label{tab:bitrate}
\resizebox{6.5cm}{!}{\begin{tabular}{@{}ll@{}}
\toprule
Resolution                 & Bitrate (Mbps)  \\\midrule
1280×720 (720p/HD) & 1.0, 1.5, 2.0, 2.5, 3.0, 3.5, 4.0, 4.5, 5.0\\
1920×1080 (1080p/FHD) & 2.0, 3.0, 4.0, 5.0, 6.0, 7.0, 8.0, 9.0, 10 \\
2560x1440 (1440p/QHD) & 4.0, 6.0, 8.0, 10, 12, 14, 16, 18, 20\\
3840x2160 (2160p/4K) & 10, 15, 20, 25, 30, 35, 40, 45, 50\\
7680x4320 (4320p/8K) & 20, 30, 40, 50, 60, 70, 80, 90, 100\\
\bottomrule
\end{tabular}}
\vspace{-3mm}
\end{table}

\vspace{-1.5mm}
\setstretch{0.96}
\subsection{Dataset and Resolution Scaling}

\begin{figure}[t!]
\centering\includesvg[width=0.9\linewidth,inkscapelatex=false]{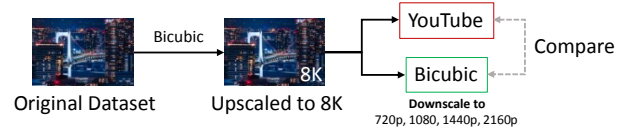}
\vspace{-1mm}
\caption{Pipeline used to obtain YouTube transcoding quality.}
\label{fig:Pipeline_YT}
\vspace{-6mm}
\end{figure}

For RD performance evaluation, two datasets were used for resolution of 4K and lower; 4K sequences of The Institute of Image Information and Television Engineers (ITE)'s Ultra-high definition/wide-color-gamut standard test sequences – Series A \cite{ITE_2016}, and Twitch dataset \cite{xiph.org}. ITE dataset consists of real-world video sequences recorded by using UHD cameras, while the Twitch dataset consists of gameplay screen recordings of the most popular games streamed by Twitch users. As for 8K resolution, only 8K sequences of the ITE dataset were used. All of the dataset has a temporal resolution of 60 fps. Since the Twitch dataset only has a resolution of 1080p, it was up-converted to 4K using Lanczos re-sampling before use in the evaluation. This is done to reflect the practice done by video game streamers, which usually run their game at the resolution below the encoding resolution, then use the Lanczos filter built into Open Broadcasting Software (OBS), one of the most popular RTMP streaming clients, to upscale the video to the final encoding resolution. Due to storage limitations of smartphones used in the experiment, raw YUV video couldn't be stored on the device, so all of the videos were encoded into H.265/HEVC using libx265 at medium preset in constant quality mode with a target CRF of 10, 8-bit main profile, chroma subsampling of 4:2:0 and inter-frame compression disabled. All downscaling was performed using bicubic re-sampling to match the transcoding pipeline employed by popular live-streaming platforms. Since only Intel QuickSync supports VP9 encoding, but the RD performance is relatively poor compared to software encoders, it was excluded from the evaluation. However, the RD curve can be seen in Figure \ref{fig:VP9Twitch4K}.

The RD performance of YouTube's VP9 and AV1 is also an interest of this study. However, YouTube only encodes non-popular videos with 8K resolution to VP9 and AV1 format, all of the videos from none 8K dataset were upscaled using bicubic re-sampling to 8K before uploading to YouTube. Note that bicubic re-sampling was used instead of the lanzcos re-sampling to match YouTube's downscaling mechanism and the upscaled 8K is not being used for the study, as this is done to force YouTube to encode the video using VP9 and AV1 codecs only. To eliminate the quality degradation from upscaling and re-encoding, the video encoded by YouTube at each resolution was compared to the lossless downscaled version of the exact video uploaded to YouTube (the bicubic upscaled 8K version). The whole process is visualized in \cref{fig:Pipeline_YT}. \looseness=-1

\begin{table}[!tbp]
\setstretch{0.9}
\caption{Comparison of average bitrate to achieve VMAF of 86-99, in step of one, between each content relative to the Twitch dataset when encoded with Intel QuickSync on Arc A770 GPU to AV1 codec.}
\vspace{-2mm}
\centering
\label{tab:VContent}
\resizebox{8.5cm}{!}{\begin{tabular}{@{}lllccc@{}}
\toprule
\multirow{2}{*}{Dataset} & \multirow{2}{*}{Source}  & \multirow{2}{*}{Resolution}  & \multirow{2}{*}{FPS} & \multirow{2}{*}{HDR} & Bitrate \\
               &  &  & &   & Diff. (\%)\\\midrule
2D Game Cutscene & Game File (VP9 10 Mbps) & 1080p/FHD & 30 & No & -75.14\\
3D Game Cutscene & Game File (VP9 10 Mbps) & 1080p/FHD & 30 & No & -67.93\\
Japanese Animation & Blu-Ray (H.264/AVC 40 Mbps) & 1080p/FHD & 24 & No & -68.11\\
VTuber Free Talk & YouTube Format 617 (H.264/AVC) & 1080p/FHD & 60 & No & -74.07\\
VTuber PV & YouTube Format 616 (VP9 Premium) & 1080p/FHD & 24 & No & -77.81\\
\midrule
3D Animation & UHD Blu-Ray (H.265/HEVC 100 Mbps) & 2160p/UHD & 24 & Yes & -80.46\\
Documentary & UHD Blu-Ray (H.265/HEVC 100 Mbps) & 2160p/UHD & 24 & Yes & -75.52\\
Film & UHD Blu-Ray (H.265/HEVC 100 Mbps) & 2160p/UHD & 24 & Yes & -20.57\\
Soccer & Satellite TV (H.265/HEVC 20 Mbps) & 2160p/UHD & 50 & Yes & -42.54\\
VLOG & YouTube Format 315 (VP9) & 2160p/UHD & 60 & No & -66.05\\

\bottomrule
\end{tabular}}
\vspace{-3mm}
\end{table}

\begin{figure}[t!]
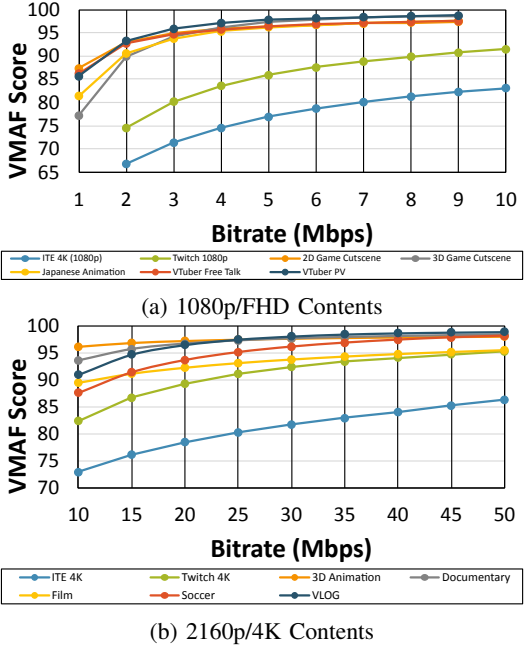

\centering
\begin{subfigure}{0.80\linewidth}
 \centering\includesvg[width=0.98\linewidth,inkscapelatex=false]{Content1080p.svg}
  \caption{1080p/FHD Contents}
  \label{fig:2KContents}
  \vspace{-1.5mm}
\end{subfigure}\\
\vspace{1mm}
\begin{subfigure}{0.80\linewidth}
 \centering\includesvg[width=0.98\linewidth,inkscapelatex=false]{Content4K.svg}
  \caption{2160p/4K Contents}
  \label{fig:4KContents}
  \vspace{-1.5mm}
\end{subfigure}

\caption{RD Curve of various contents when being encoded by Intel QuickSync on Arc A770 GPU to AV1 codec.}
\vspace{-5mm}
\label{fig:ContentRD}
\end{figure}

\subsection{Objective Evaluation Metrics}

As for objective quality evaluation metrics, three objective evaluation metrics, Peak Signal-to-Noise Ratio (PSNR), Structural Similarity Index Measure (SSIM), as well as Video Multi-Method Assessment Fusion (VMAF), a machine-learning-based quality metric, were used. VMAF was recently proposed by Netflix to address the issue that high PSNR doesn't necessarily yield high subjective quality \cite{vmaf}. The standard model with the version of 0.6.1 was used for a resolution of 4K and below, while the 4K variant with the version of 0.6.1 was used for 8K, then the average VMAF score across all video frames was taken for each video sequence. The standard model was used for 4K resolution and below because these days many devices have a high pixel density display, so 4K content consumption is not limited to TV anymore, whereas consumption of 8K content is still mostly limited to large TV screens only, so using the VMAF 4K model for 8K resolution is appropriate. By using the data obtained from the experiment, a logarithm best-fit curve was calculated for each test case, and then compared to each other. RD performance comparison is being split into three different bitrate ranges: low bitrate, medium bitrate, and high bitrate. By using the logarithm best fit, five data points equally spaced were taken from low and high ranges, while ten data points were obtained from the medium bitrate range. Then the average was taken across the data points in each range. The range used for each resolution can be found in \cref{tab:bitrateRange}.

\begin{table}[!tbp]
\setstretch{0.85}
\caption{Bitrate range used for the evaluations}
\vspace{-1.5mm}
\centering
\label{tab:bitrateRange}
\resizebox{6.5cm}{!}{\begin{tabular}{@{}lccc@{}}
\toprule
Resolution                 & Low & Medium & High  \\\midrule
1280×720 (720p/HD) & 1.0-2.0 & 2.0-4.0 & 4.0-5.0\\
1920×1080 (1080p/FHD) & 2.0-4.0 & 4.0-8.0 & 8.0-10 \\
2560x1440 (1440p/QHD) & 4.0-8.0 & 8.0-16 & 16-20\\
3840x2160 (2160p/4K) & 10-20 & 20-40 & 40-50\\
7680x4320 (4320p/8K) & 20-40 & 40-80 & 80-100\\
\bottomrule
\end{tabular}}
\vspace{-5mm}
\end{table}


\looseness=-1


\subsection{Comparison of the Dataset to Real-world contents}
Realizing that the dataset used for the evaluation contains extremely complex scenes intended to push the encoders to their limit, which may not be a good representation of everyday content, ten different types of commonly found content from both high-quality physical media sources and online distribution were encoded using AV1 codec on Intel Arc A770 GPU, then the RD curve was plotted and compared to both ITE 4K and Twitch dataset. Each of the content types contains 6-22 videos. The description and source format information can be found in \cref{tab:VContent}, and the RD curve can be seen in \cref{fig:ContentRD}. Note that HDR content will be encoded to HDR output without down-converting to SDR. There were five 1080p content sets. 2D/3D Game Cutscenes were taken from Genshin Impact. Japanese animations consisted of intro and outro parts of various titles ripped from commercial Blu-Ray discs. VTuber Free Talk consisted of a randomly cut 30-second-long segment of every video in "ARParty : Cancer Awareness" playlist. Finally, VTuber PV consisted of 11 PV from Algorthythm Project (ARP) that had gained enough popularity for YouTube to provide \textit{Enhanced Bitrate} quality. As for five UHD content sets, Animation and Film were ripped from UHD Blu-Ray of various titles. The documentary was ripped from UHD Blu-Ray of BBC's Planet Earth II. Soccer is the final match of FIFA World Cup 2018 recorded from a 4K satellite broadcast. Lastly, VLOG consisted of randomly cut segments of "Jet Lag: The Game," a travel web series published on YouTube.

From the results, it can be said that the dataset used in this paper is relatively difficult to compress compared to other real-world content. \cref{fig:ContentRD} shows that most real-world contents can reach a VMAF score of 90 despite using way less bitrate compared to our dataset. While this suggested that our dataset is capable of pushing the encoder's capability, in reality, lower bitrate may be used to save the bandwidth as most of the contents may not require a very high bitrate to achieve good quality. \looseness=-1


\section{Results and Analysis}

\subsection{Encoding Throughput}
\looseness=-1
To determine the software encoder preset that can realistically be used for real-time encoding, a 1080p60 video was transcoded on a laptop with a high-end CPU. Hardware decoders were used to ensure that the video decoding process didn't put an additional load on the CPU. The preset was increased until it couldn't keep up with real-time encoding on the given hardware, which is when the encoding throughput dropped below 60 fps. It was found that the best presets that can be handled by Intel Core i7-13700H were slow for libx264, faster for libx265, and preset 8 for SVT-AV1, which is when the CPU power consumption hits the limit in Table \ref{tab:EncodingSpeed1080p}. Therefore, these presets will be used to represent software encoders. When looking at the encoding speed at 4K in \cref{tab:EncodingSpeed4K}, it was found that the encoding frame rate of the latest GPU boards can easily exceed the very fast preset on high-end CPUs with some GPU boards able to even exceed the speed of the fastest preset in some codec.  However, it should be noted that these results are based on the hardware available at the time of research, and advancement in microprocessors technology might allow slower presets to become feasible for real-time encoding in the future.

\subsection{Power Consumption}

\looseness=-1
During the experiment, the CPU utilization and total system power consumption were also recorded for comparison to hardware encoders. It was found that when using the software encoders with the best possible preset, the total system power consumption soared to double the amount used by the hardware encoders. Since the system CPU is being heavily utilized, less headroom is available for other tasks, which may cause stuttering to the content that the user is trying to broadcast, especially video games running on the same computer, and lead to undesirable effects of stuttering and frame drop. When considering the encoding frame rate at 4K on a high-end desktop CPU (see Table \ref{tab:EncodingSpeed4K}), it was found that hardware encoders, including smartphone SoCs, can easily outperform the encoding speed of the faster presets on the software encoders, especially when using newer codecs like H.265/HEVC and AV1. Considering that the CPU used for this experiment was a high-end 105W Ryzen 9 5900X, yet it yielded a lower encoding frame rate than hardware counterparts when using reasonably high-quality presets, demonstrates that hardware encoders were able to deliver much higher energy efficiency at video encoding tasks.

\begin{table}[!tbp]
\setstretch{0.85}
\caption{System Resource Utilization when perform Real-Time Encoding on Intel Core i7-13700H targeting 1080p60}
\vspace{-1.5mm}
\centering
\label{tab:EncodingSpeed1080p}
\resizebox{8cm}{!}{\begin{tabular}{@{}lcccccc@{}}
\toprule
\multirow{2.5}{*}{Encoder}    & \multicolumn{2}{c}{AVC} & \multicolumn{2}{c}{HEVC} & \multicolumn{2}{c}{AV1} \\\cmidrule(lr){2-3}\cmidrule(lr){4-5}\cmidrule(l){6-7}
 & CPU (\%) & Pwr (W) &CPU (\%) & Pwr (W)&CPU (\%) & Pwr (W)\\\midrule
Intel 13th Generation & 5.2 & 28.23 & 5.4 & 29.42 & N/A & N/A\\
NVIDIA Ada Lovelace & 4.9 & 37.35 & 4.7 & 40.38 & 4.8 & 37.58\\\midrule
Software Slow & 38.3 & 85.27 & N/A & N/A & N/A & N/A\\
Software Medium/P6 & 20.8 & 77.41 & N/A & N/A & N/A & N/A\\
Software Fast/P7 & 19.9 & 75.14 & N/A & N/A & N/A & N/A\\
Software Faster/P8 & 16.4 & 68.82 & 43.4 & 93.48 & 67.4 & 102.38\\
Software Very Fast/P9 & 12.6 & 56.85 & 41.2 & 92.10 & 35.9 & 88.66\\
Software Super Fast/P10 & 7.7 & 42.50 & 38.5 & 86.74 & 24.8 & 72.78\\
Software Ultra Fast/P11 & 7.6 & 33.70 & 29.7 & 35.83 & 18.8 & 61.54\\

\bottomrule
\end{tabular}}
\vspace{-2.5mm}
\end{table}

\begin{table}[!tbp]
\setstretch{0.85}
\caption{Encoding Throughput at 2160p/4K (Frames per second). Software Encoders was ran on Ryzen 9 5900X CPU.}
\vspace{-1.5mm}
\centering
\label{tab:EncodingSpeed4K}
\resizebox{6cm}{!}{\begin{tabular}{@{}lccc@{}}
\toprule
Encoder    & AVC & HEVC & AV1 \\\midrule
Qualcomm Snapdragon 888 & 43.16 & 48.55 & N/A\\
Qualcomm Snapdragon 8 Gen 1 & 47.05 & 46.15 & N/A\\
Qualcomm Snapdragon 8 Gen 2 & 56.34 & 63.24 & N/A\\\midrule
Intel 7th Generation & 36.86 & 13.46 & N/A\\
Intel 11th Generation & 24.97 & 13.46 & N/A\\
Intel 13th Generation & 51.55 & 21.37 & N/A\\
Intel Arc GPU & 62.04 & 55.14 & 47.65\\\midrule
NVIDIA Maxwell 1st Generation & 17.86 & N/A & N/A \\
NVIDIA Maxwell 2nd Generation & 34.17 & 17.44 & N/A \\
NVIDIA Pascal & 53.95 & 28.65 & N/A \\
NVIDIA Turing & 56.34 & 29.85 & N/A \\
NVIDIA Ampere & 56.34 & 29.34 & N/A \\
NVIDIA Ada Lovelace & 69.53 & 37.16 & 64.74 \\\midrule
Software Medium/P6 & 29.28 & 14.23 & 5.81\\
Software Fast/P7 & 30.57 & 24.57 & 12.35\\
Software Faster/P8 & 35.36 & 26.88 & 23.86\\
Software Very Fast/P9 & 39.86 & 26.90 & 31.47\\
Software Super Fast/P10 & 104.90 & 31.44 & 39.26\\
Software Ultra Fast/P11 & 157.04 & 39.87 & 53.65\\

\bottomrule
\end{tabular}}
\vspace{-6.5mm}
\end{table}

\subsection{Rate-Distortion Performance}

In the previous sub-section, it was found that hardware encoders can provide higher encoding speed, while consuming less power, but rate-distortion performance is argueable one of the most important aspects. It's known that at slower presets, software encoders yield a higher video quality than hardware encoders at the same bitrate. However, these slower presets are very CPU intensive and couldn't keep up with real-time encoding tasks for live-streaming. Therefore, it wouldn't be fair to compare hardware encoders to the slowest software encoder preset. Instead, the slowest presets that can yield real-time encoding at 1080p60 from the previous subsection will be used. \looseness=-1

From the experiment, when considering the VMAF score (see \cref{tab:BitrateComparedVMAF} and \cref{tab:BitrateRelativeVMAF}), it was found that at 1080p/FHD, Intel QuickSync yields an average of 0.51 higher VMAF score compared to the software encoder, compared to an advantage of 0.38 gained by NVIDIA encoder. However, Qualcomm's implementation didn't support B-Frame encoding, so the compression efficiency took a big hit resulting in a 6.77 less VMAF score compared to the software encoder. This effect is clearly shown at the low bitrate range as the Qualcomm encoder falls behind the software encoder by around 10.20 points, which is very significant. All of the encoders perform better when the bitrate is being increased. Intel encoders yield 0.11 point advantage at low bitrate, but the advantage grew to 0.85 at high bitrate. This is also true for NVIDIA, which has its advantage increased from 0.19 at low bitrate to 0.54 at high bitrate range. Interestingly, at 1080p, the NVIDIA encoder seems to perform slightly better than Intel at an extremely low bitrate, but the lead quickly evaporates when the bitrate is increased. Finally, while raising the bitrate helps Qualcomm encoder perform better, it still couldn't keep up with the software counterpart, as it still falling behind the software encoders by an average of 3.85 points at high bitrate across all video codecs. 

However, when looking from the PSNR perspective (see \cref{tab:BitrateComparedPSNR}), it was found that at 1080p, NVIDIA encoders perform slightly better than Intel, yielding an average of 0.27 dB higher PSNR than software encoder compared to 0.11 dB leads by Intel across all bitrate and codecs. Qualcomm still falls behind as it yields 1.30 dB lower PSNR compared to the software encoder. Similar to VMAF, it was observed that at higher bitrate, hardware encoders perform better either taking the advantage further or closing the gap slightly. Unfortunately, SSIM didn't paint any clear picture of the difference between NVIDIA and Intel encoders as both yield a very similar score (see \cref{tab:BitrateComparedSSIM}). However, it confirmed that the Qualcomm encoder yielded worse RD performance, losing by 0.0082 points against software encoders.

Next, at 2160p/4K resolution, it was found that all of the hardware encoders performed better compared to 1080p/FHD when considering the average relative performance to software encoders. When considering the VMAF score, Intel now yielded a 0.75 point advantage compared to 0.51 at 1080p, NVIDIA stayed relatively the same, and Qualcomm started to catch up as it only lost to the software encoder by an average of 5.30 points at this resolution. Interestingly, compared to 1080p, all of the hardware encoders gain an advantage at the low bitrate range, while losing its lead at the high bitrate range. As for PSNR, at this resolution, Intel and NVIDIA are now both comparable as they yield 0.29 dB and 0.36 dB advantage, respectively. Qualcomm is still losing by 0.82 dB, a slight improvement over the lower 1080p resolution.

\begin{figure}[t!]
\centering
\begin{subfigure}{.24\textwidth}
  \centering
\centering\includesvg[width=0.95\linewidth,inkscapelatex=false]{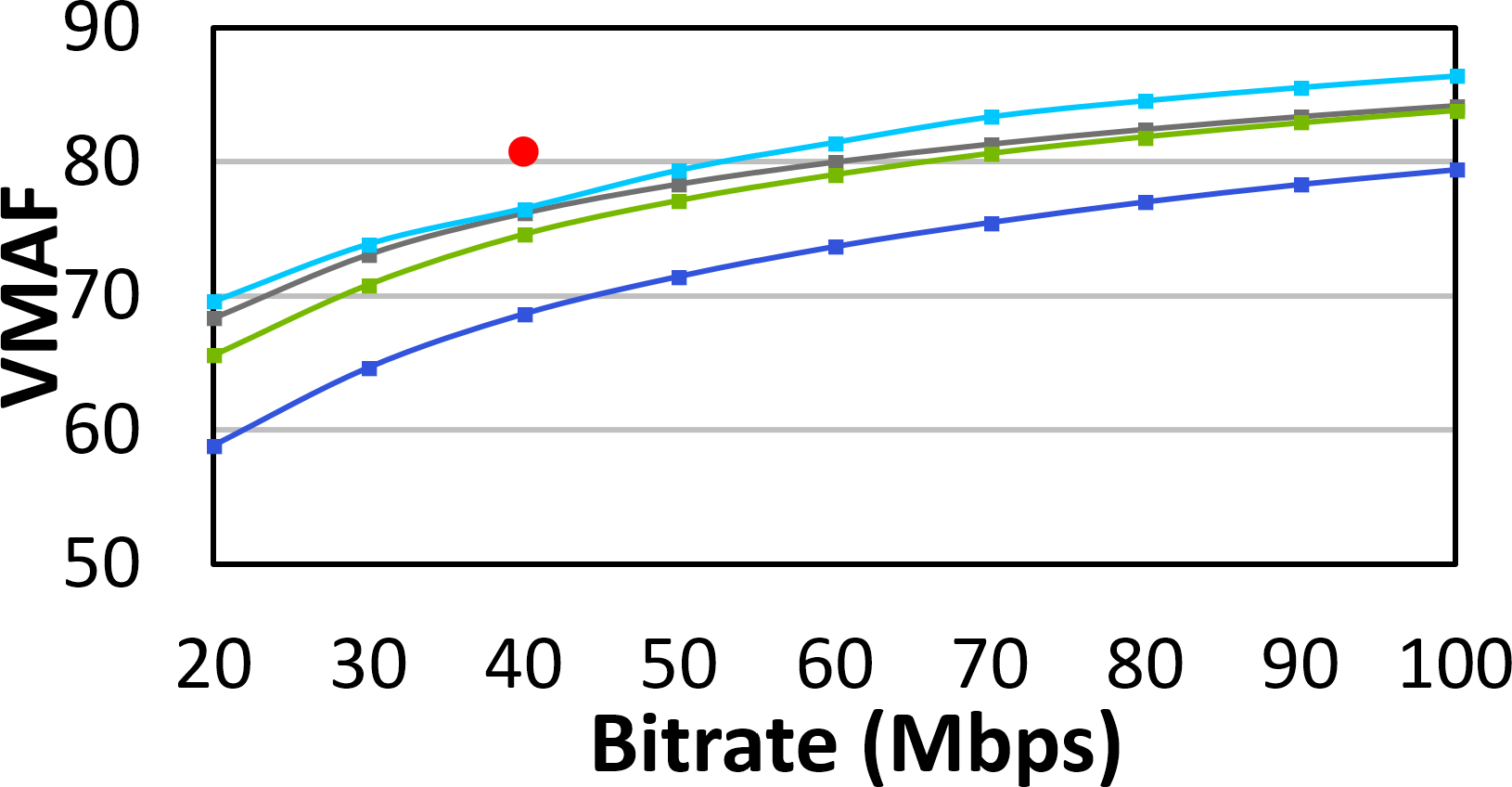}
  \vspace{-1.5mm}
  \caption{ITE 8K - HEVC 4320p/8K}
  \label{fig:HEVCITE8K}
\end{subfigure}%
\begin{subfigure}{.24\textwidth}
\centering\includesvg[width=0.95\linewidth,inkscapelatex=false]{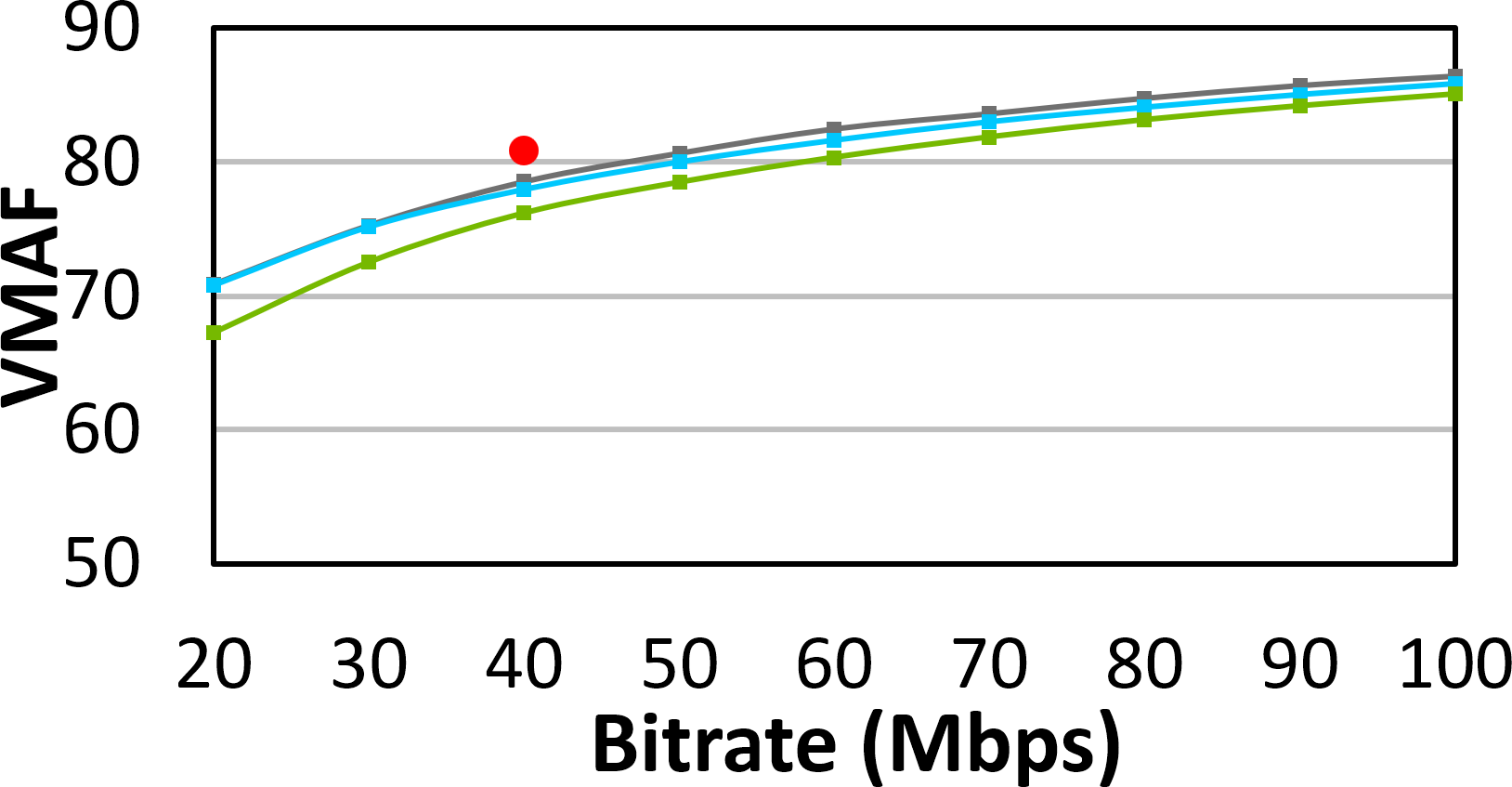}
  \vspace{-1.5mm}
  \caption{ITE 8K - AV1 4320p/8K}
  \label{fig:AV1ITE8K}
\end{subfigure}\\
\begin{subfigure}{.24\textwidth}
\centering\includesvg[width=0.95\linewidth,inkscapelatex=false]{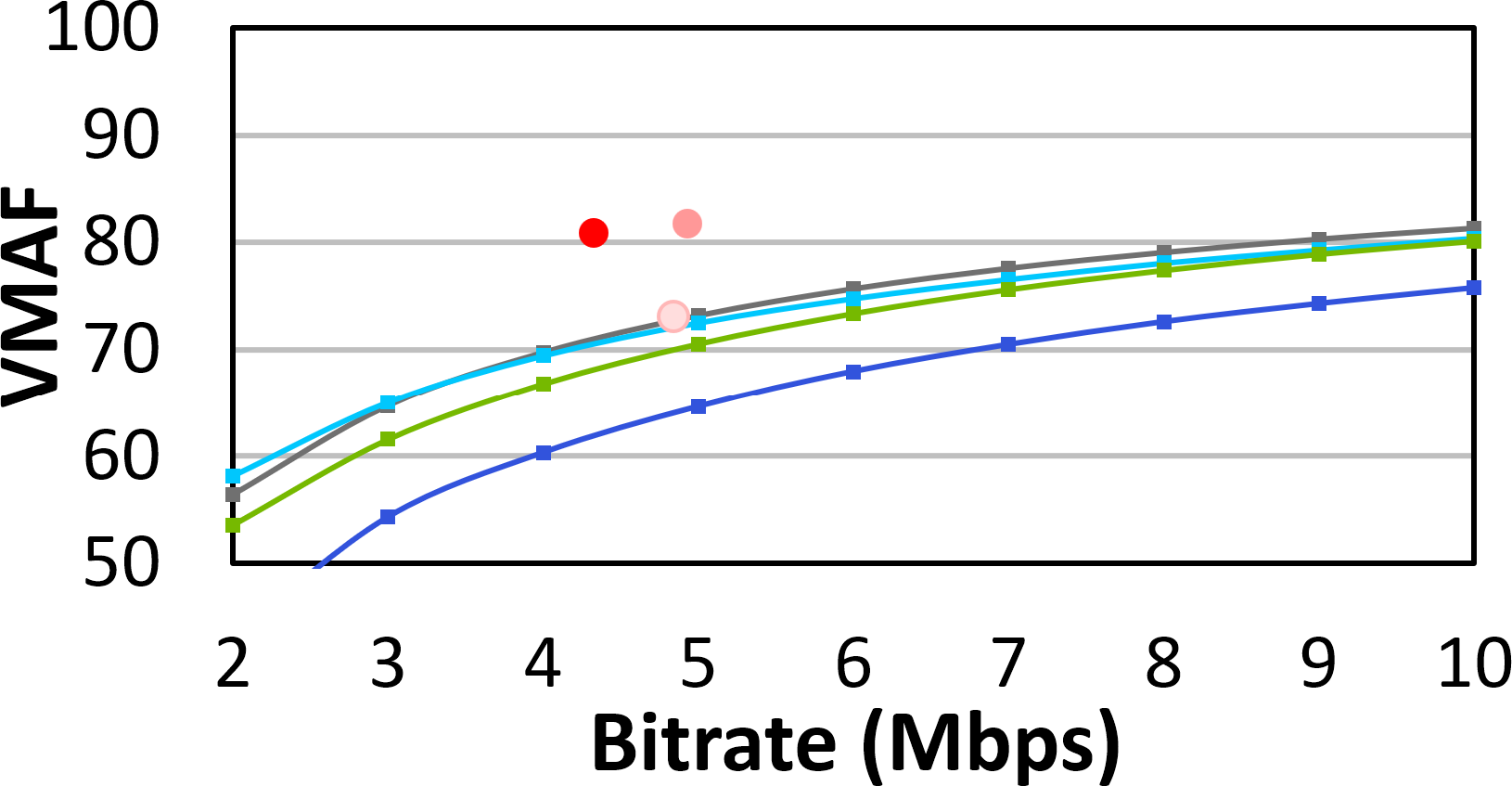}
  \vspace{-1.5mm}
  \caption{ITE 4K - AVC 1080p/FHD}
  \label{fig:AVCITE2K}
\end{subfigure}%
\begin{subfigure}{.24\textwidth}
\centering\includesvg[width=0.95\linewidth,inkscapelatex=false]{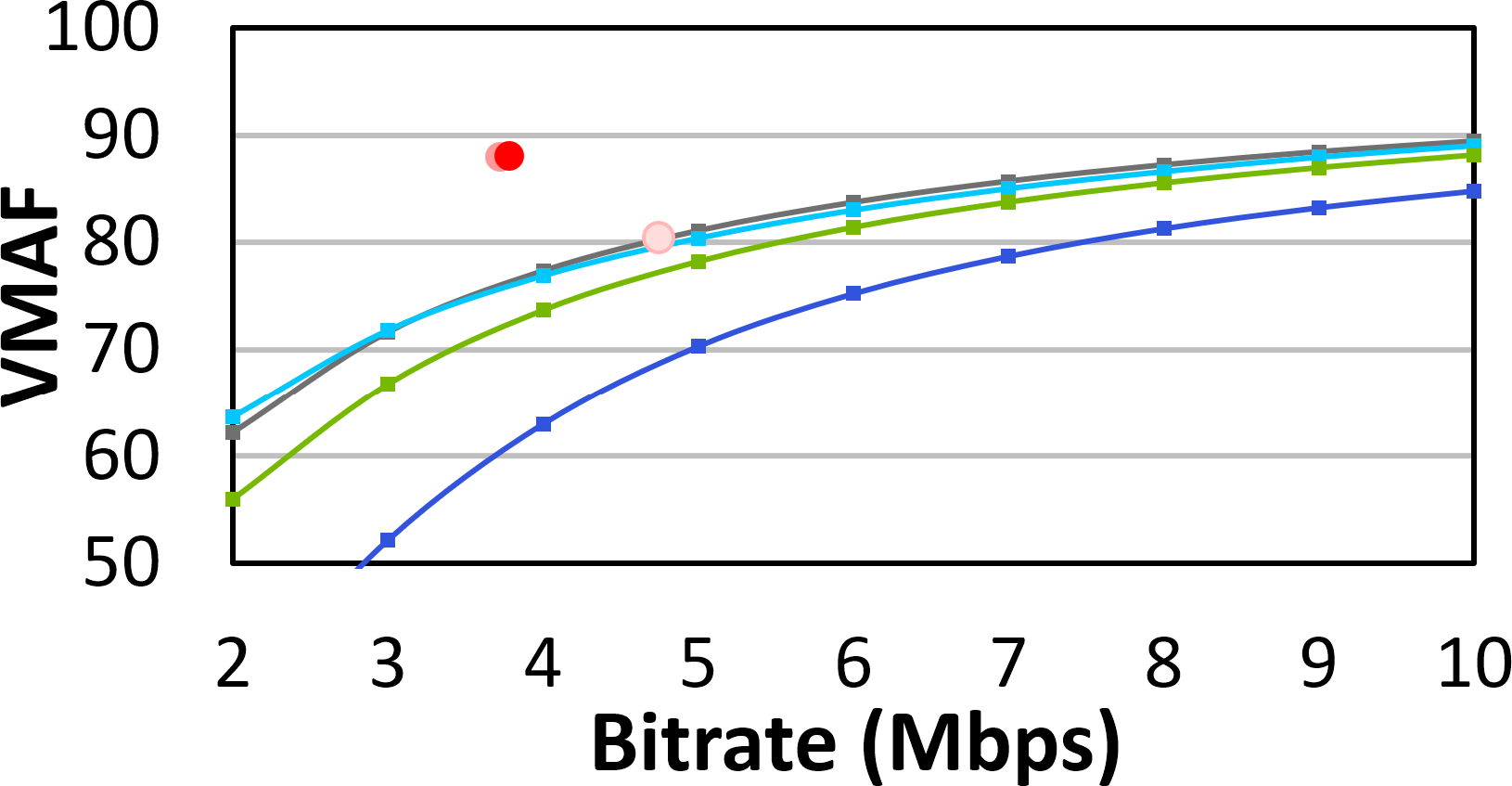}
  \vspace{-1.5mm}
  \caption{Twitch - AVC 1080p/FHD}
  \label{fig:AVCTwitch2K}
\end{subfigure}\\
\begin{subfigure}{.24\textwidth}
\centering\includesvg[width=0.95\linewidth,inkscapelatex=false]{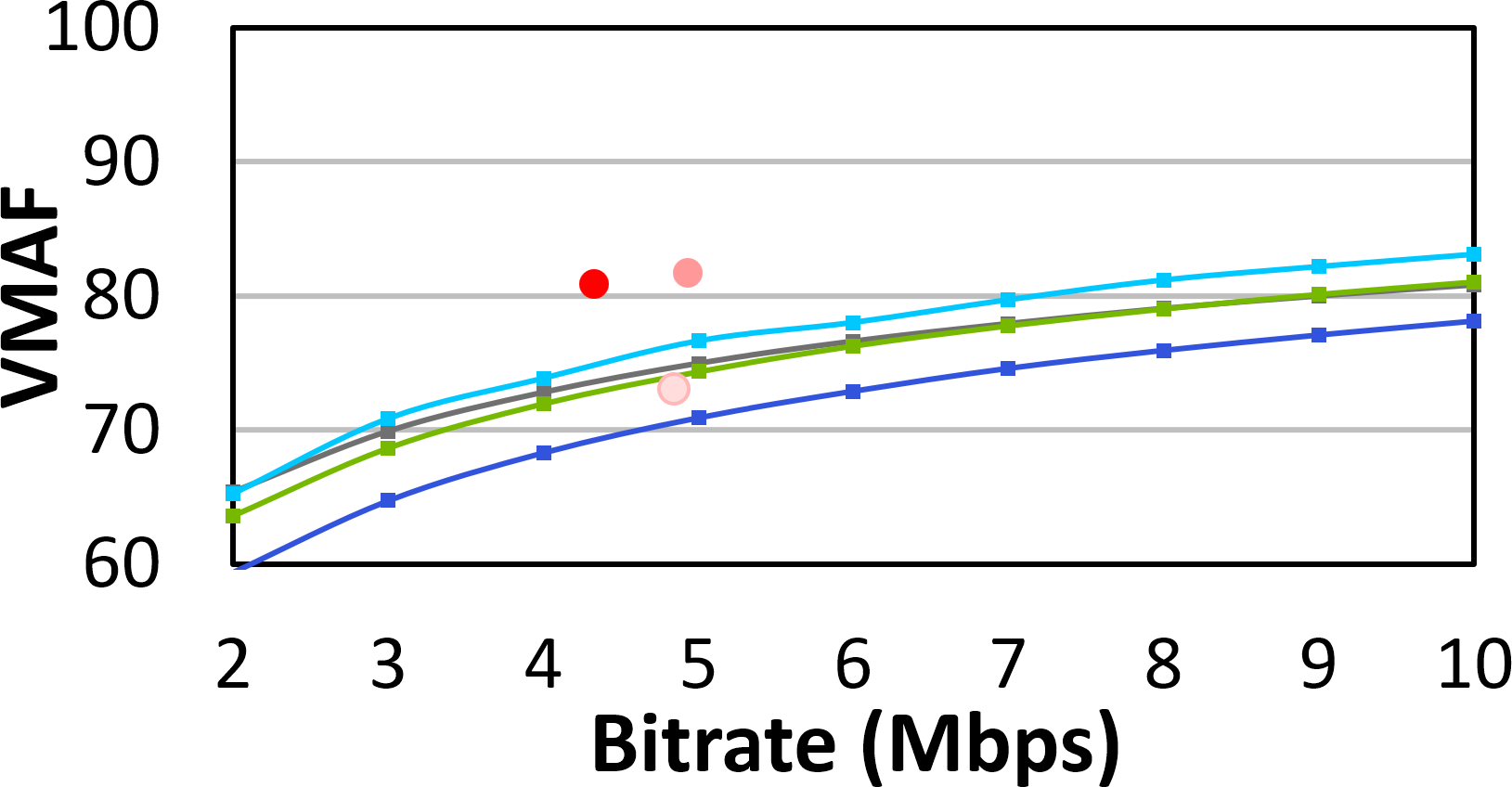}
  \vspace{-1.5mm}
  \caption{ITE 4K - HEVC 1080p/FHD}
  \label{fig:HEVCITE2K}
\end{subfigure}%
\begin{subfigure}{.24\textwidth}
\centering\includesvg[width=0.95\linewidth,inkscapelatex=false]{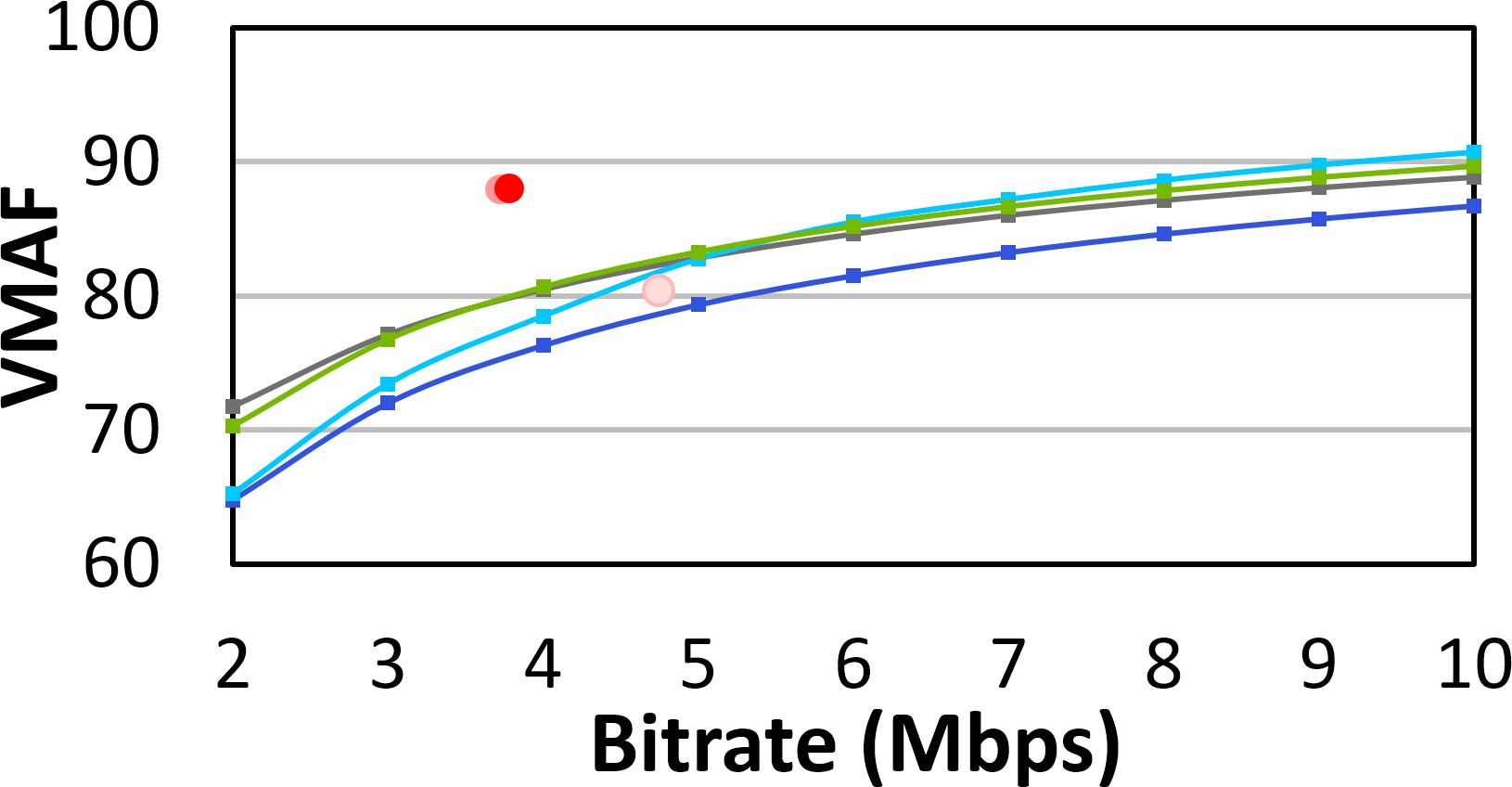}
  \vspace{-1.5mm}
  \caption{Twitch - HEVC 1080p/FHD}
  \label{fig:HEVCTwitch2K}
\end{subfigure}\\
\begin{subfigure}{.24\textwidth}
\centering\includesvg[width=0.95\linewidth,inkscapelatex=false]{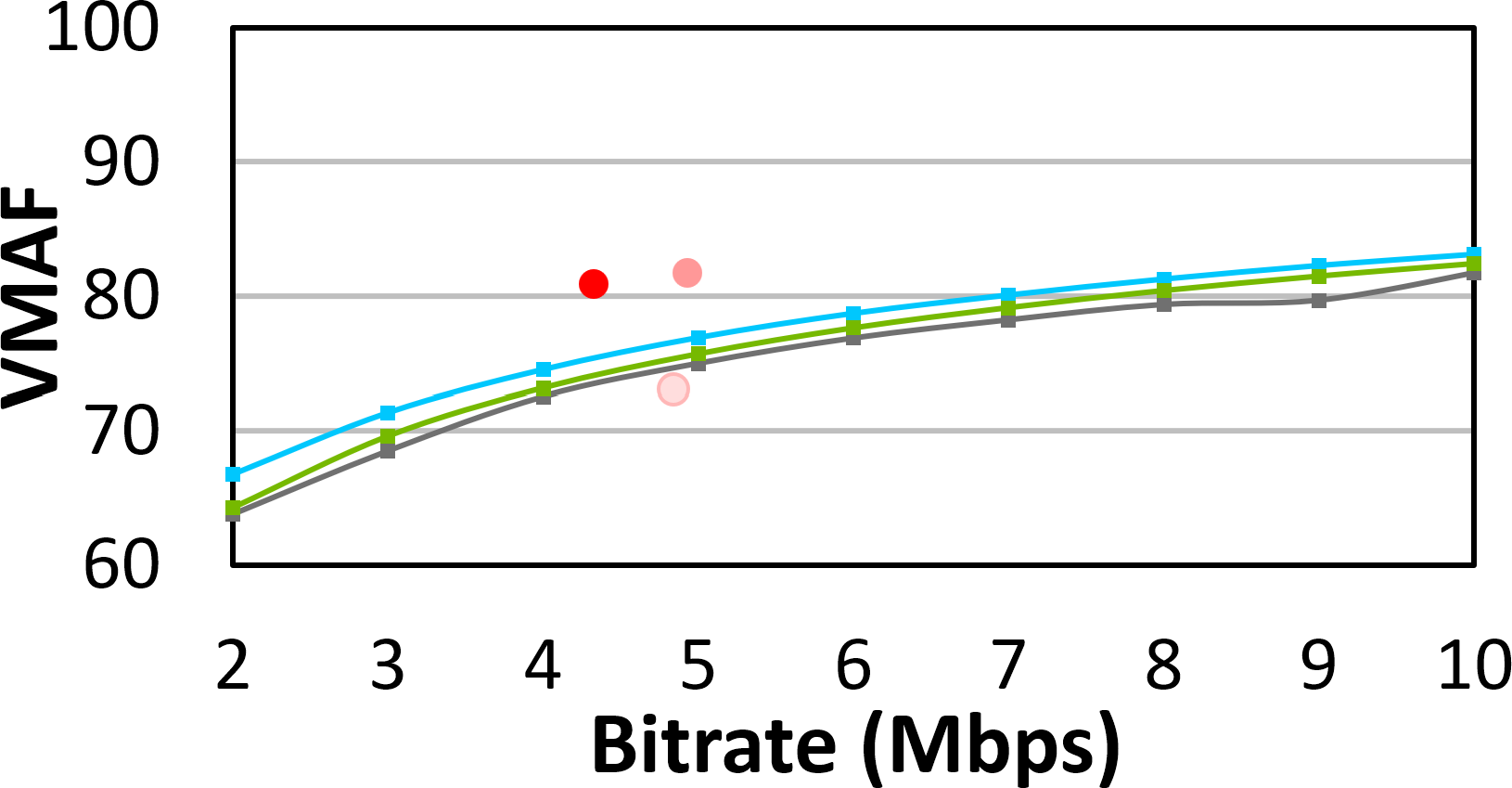}
  \vspace{-1.5mm}
  \caption{ITE 4K - AV1 1080p/FHD}
  \label{fig:AV1ITE2K}
\end{subfigure}%
\begin{subfigure}{.24\textwidth}
  \centering
\centering\includesvg[width=0.95\linewidth,inkscapelatex=false]{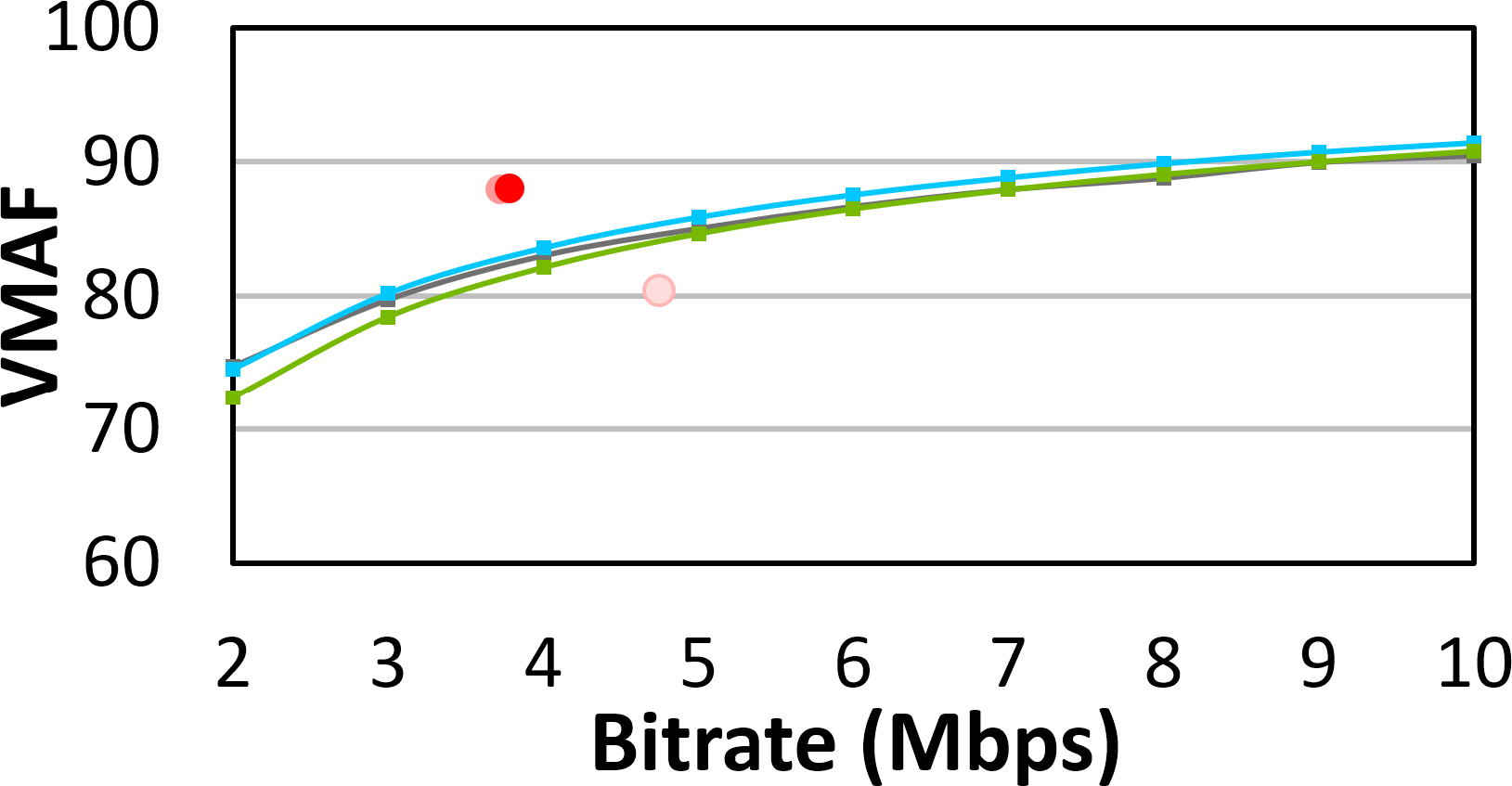}
  \vspace{-1.5mm}
  \caption{Twitch - AV1 1080p/FHD}
  \label{fig:AV1Twitch2K}
\end{subfigure}
\\
\begin{subfigure}{.24\textwidth}
  \centering
\centering\includesvg[width=0.95\linewidth,inkscapelatex=false]{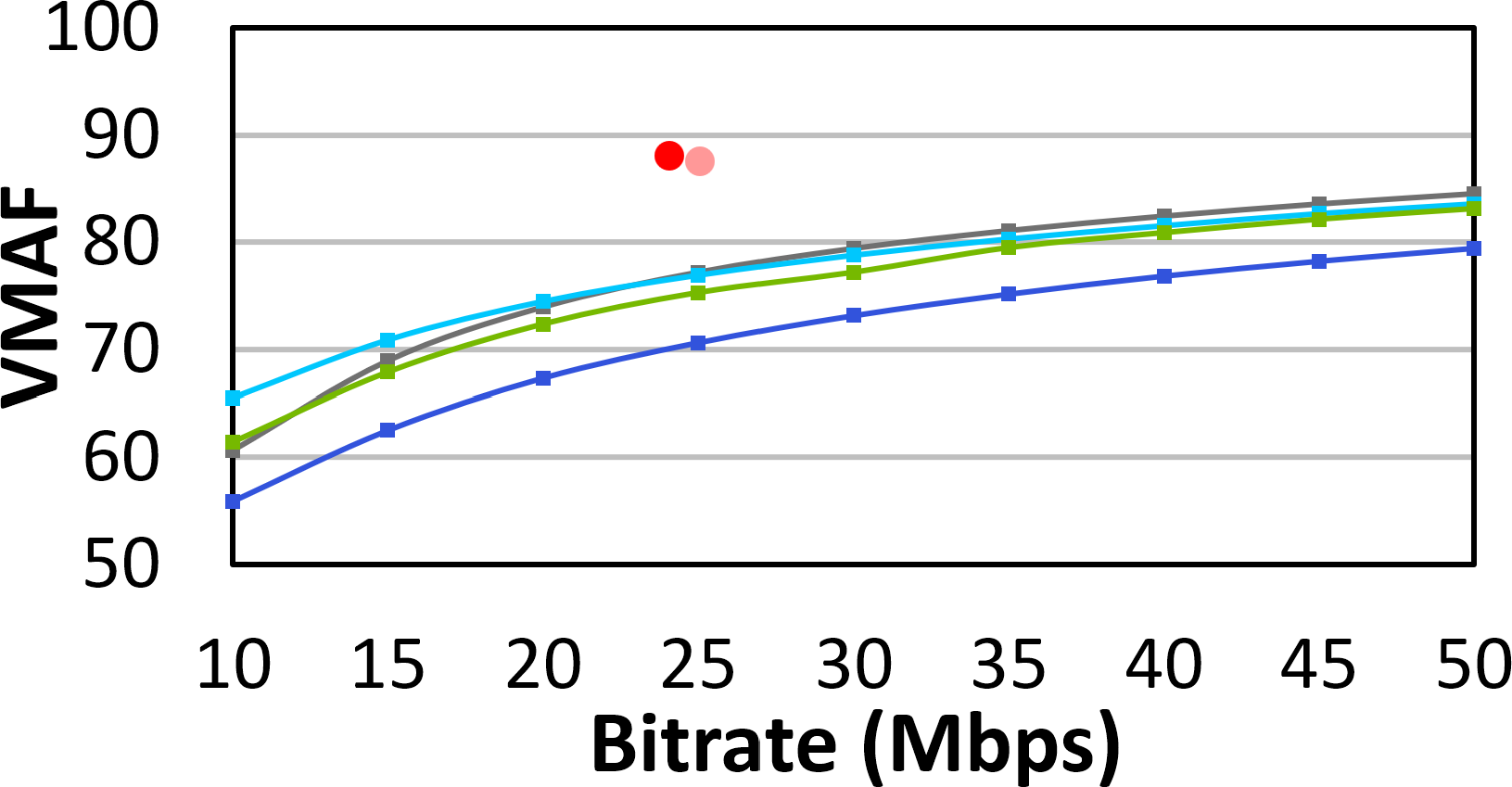}
  \vspace{-1.5mm}
  \caption{ITE 4K - AVC 2160p/4K}
  \label{fig:AVCITE4K}
\end{subfigure}%
\begin{subfigure}{.24\textwidth}
\centering\includesvg[width=0.95\linewidth,inkscapelatex=false]{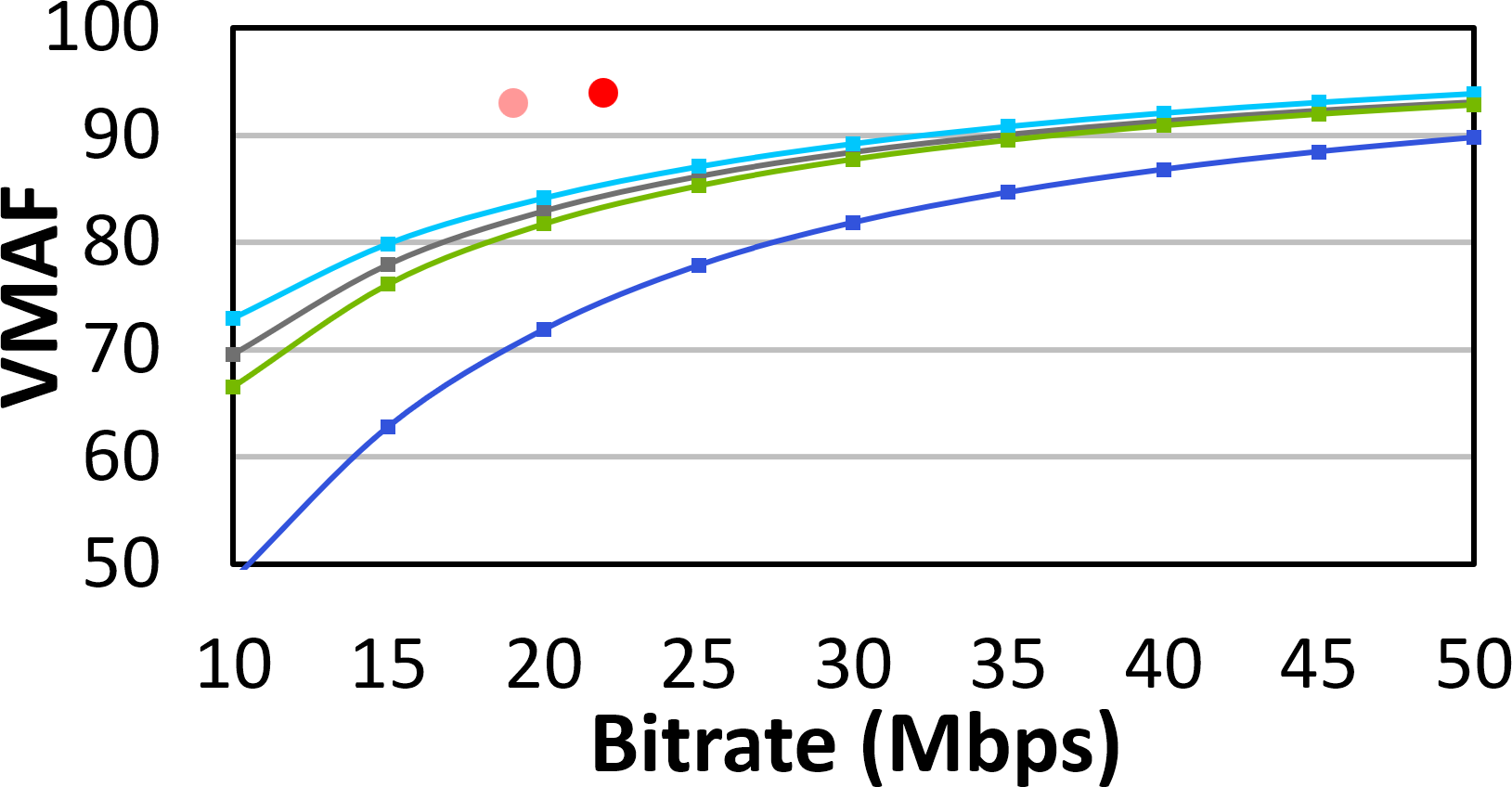}
  \vspace{-1.5mm}
  \caption{Twitch - AVC 2160p/4K}
  \label{fig:AVCTwitch4K}
\end{subfigure}\\
\begin{subfigure}{.24\textwidth}
  \centering
\centering\includesvg[width=0.95\linewidth,inkscapelatex=false]{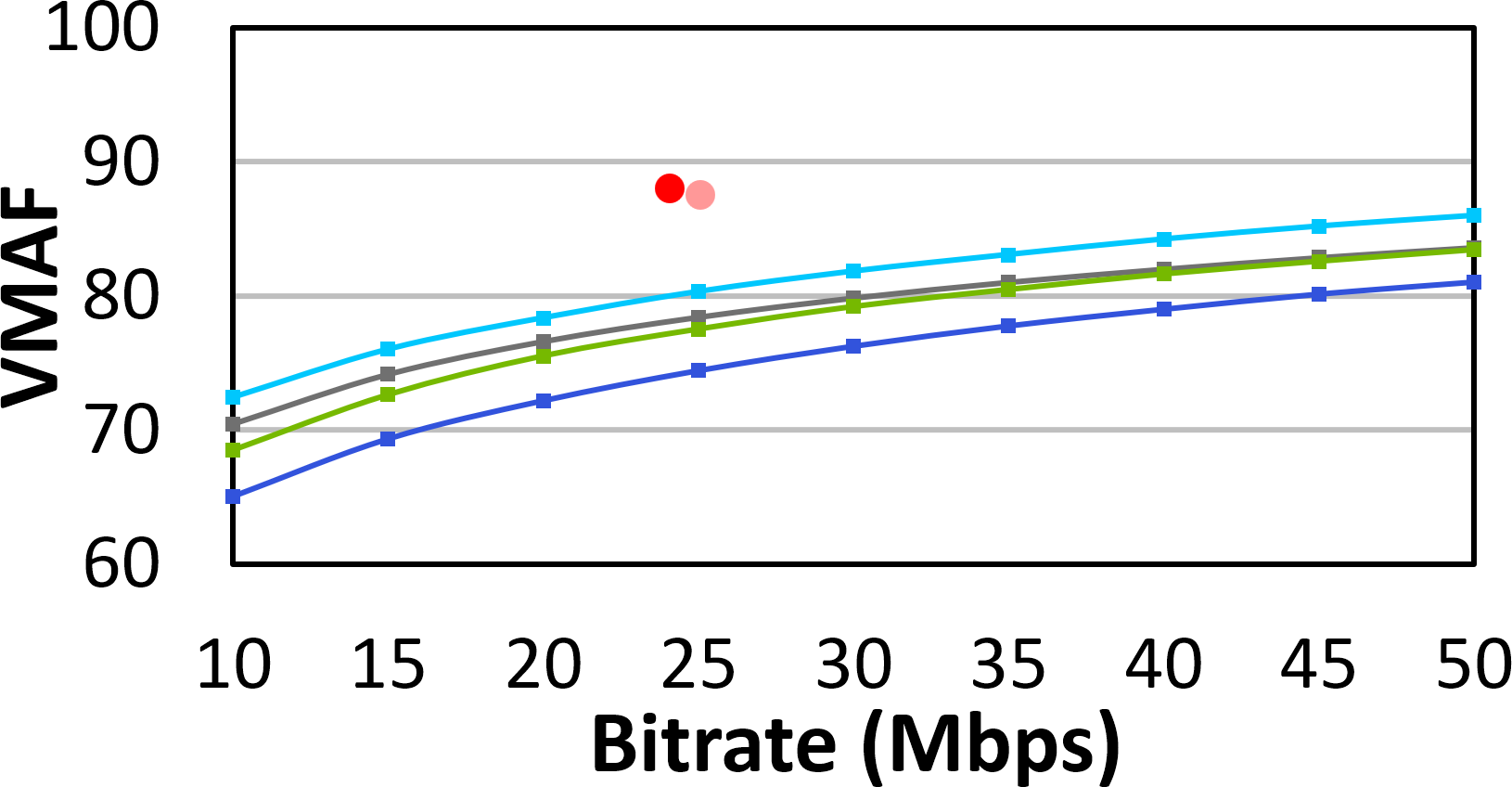}
  \vspace{-1.5mm}
  \caption{ITE 4K - HEVC 2160p/4K}
  \label{fig:HEVCITE4K}
\end{subfigure}%
\begin{subfigure}{.24\textwidth}
\centering\includesvg[width=0.95\linewidth,inkscapelatex=false]{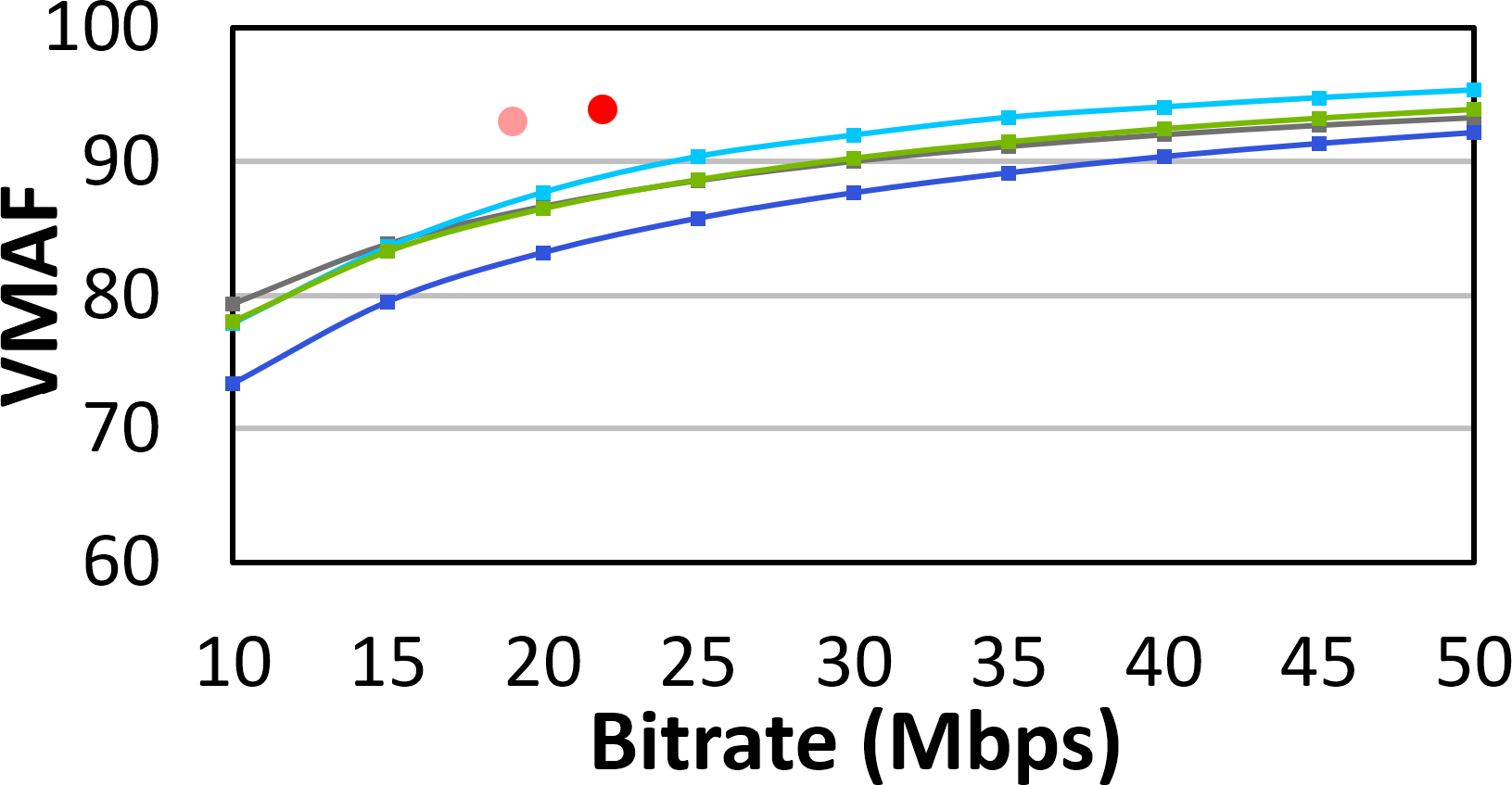}
  \vspace{-1.5mm}
  \caption{Twitch - HEVC 2160p/4K}
  \label{fig:HEVCTwitch4K}
\end{subfigure}\\
\begin{subfigure}{.24\textwidth}
  \centering
\centering\includesvg[width=0.95\linewidth,inkscapelatex=false]{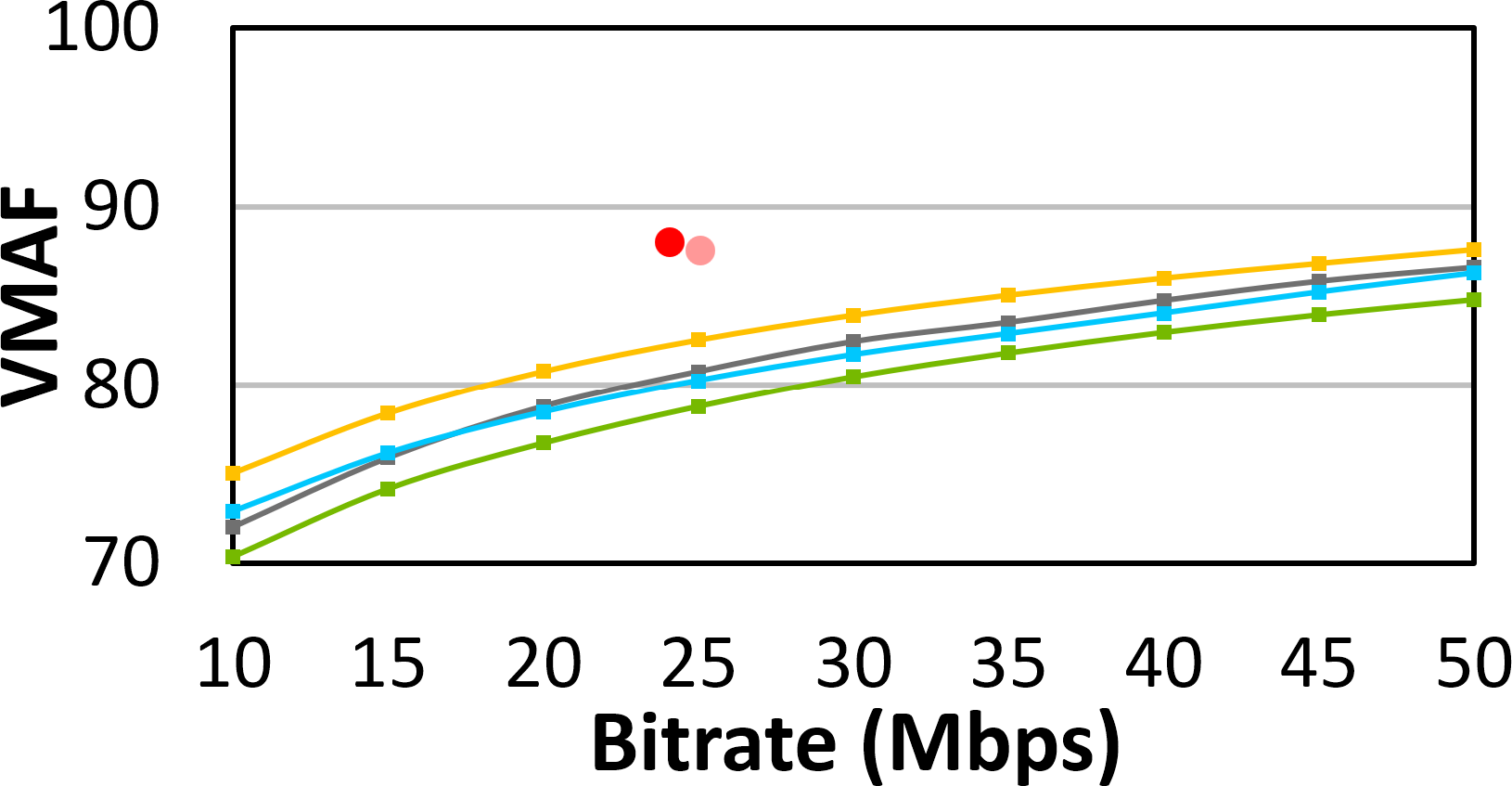}
  \vspace{-1.5mm}
  \caption{ITE 4K - AV1 2160p/4K}
  \label{fig:AV1ITE4K}
\end{subfigure}%
\begin{subfigure}{.24\textwidth}
\centering\includesvg[width=0.95\linewidth,inkscapelatex=false]{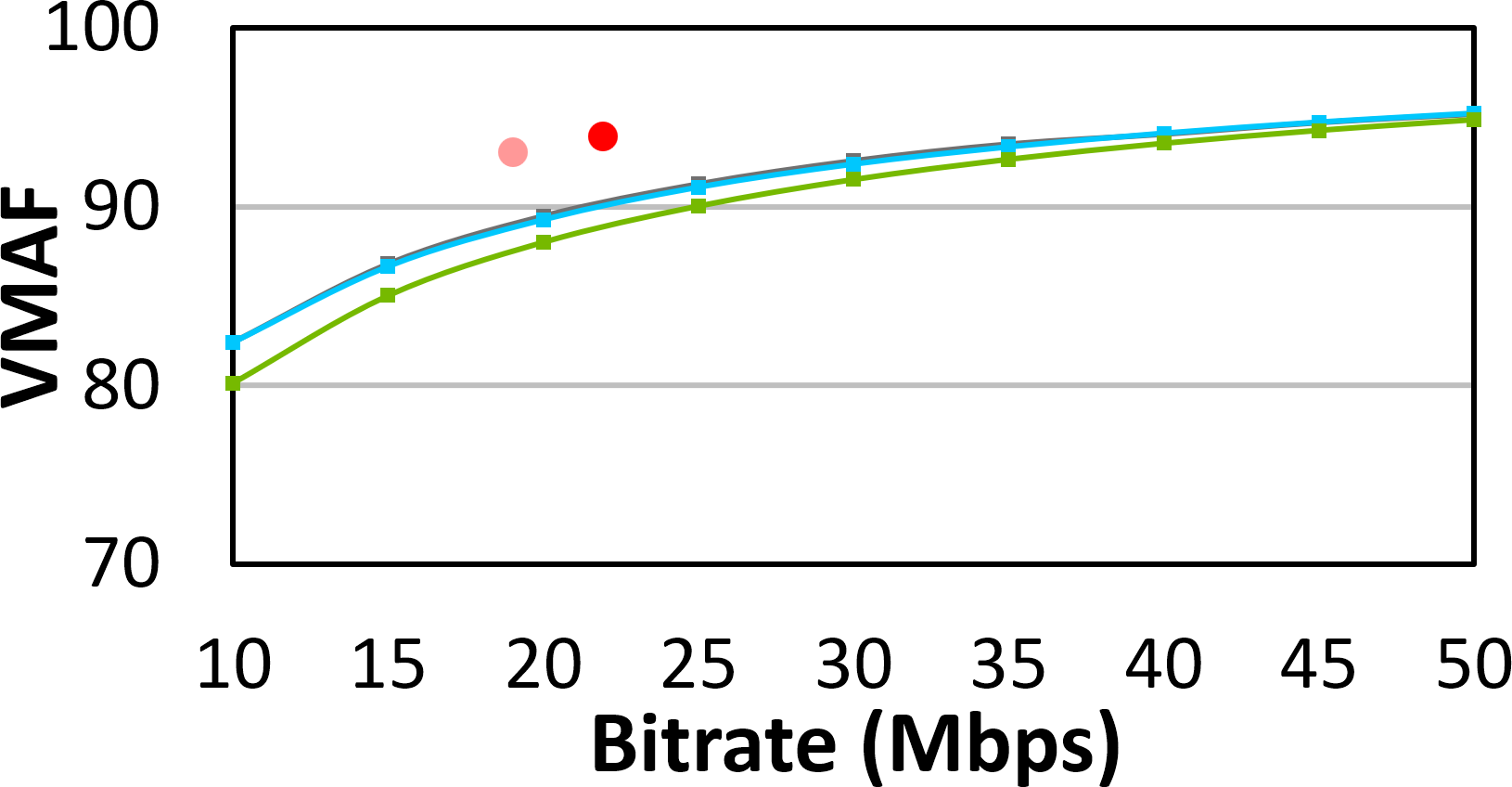}
  \vspace{-1.5mm}
  \caption{Twitch - AV1 2160p/4K}
  \label{fig:AV1Twitch4K}
\end{subfigure}\\
\begin{subfigure}[T]{.24\textwidth}
  \centering
  \vspace{-3mm}
\vspace{-1.5mm}\centering\includesvg[width=0.95\linewidth,inkscapelatex=false]{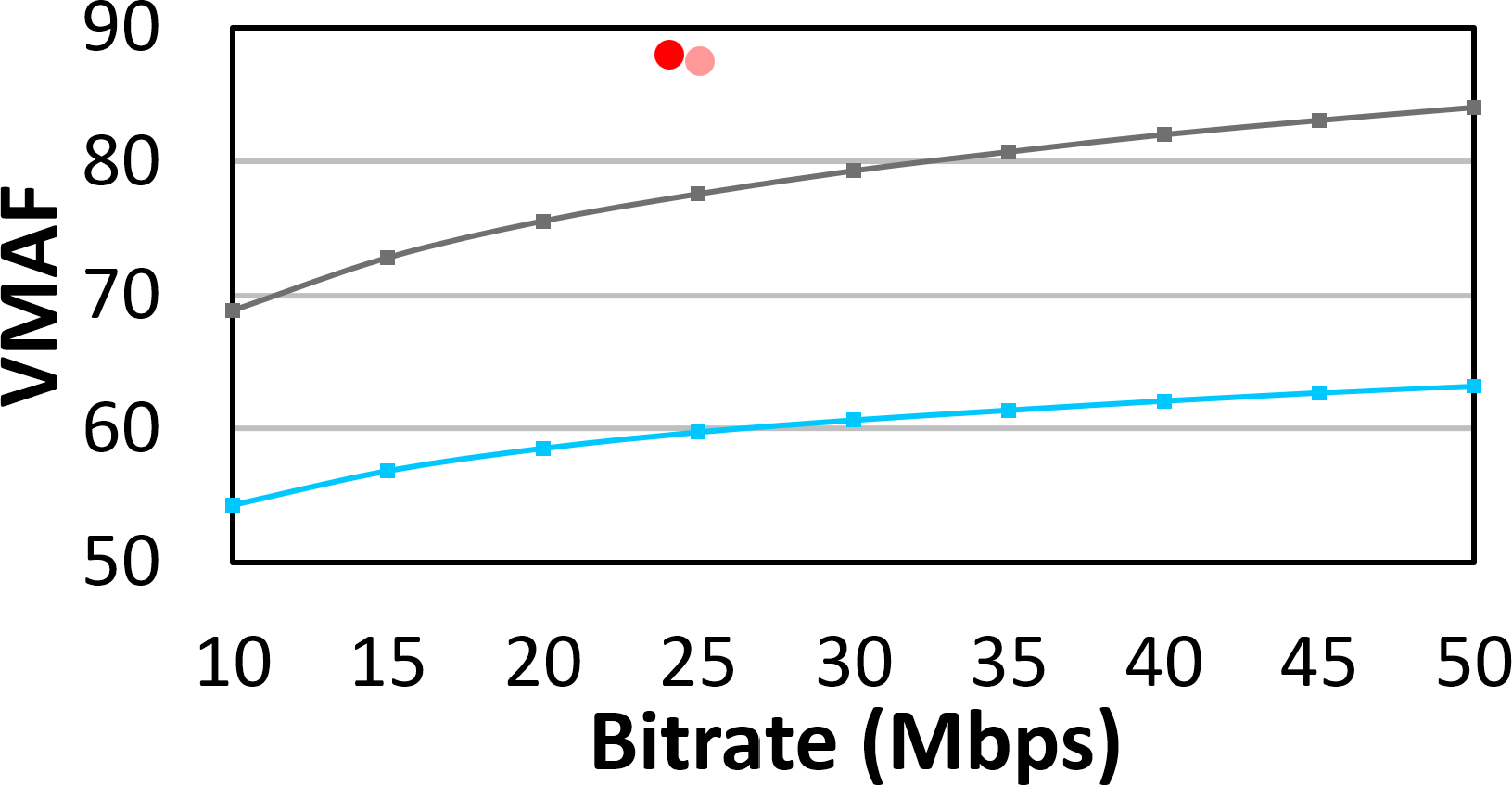}
  \vspace{-1.5mm}
  \caption{ITE 4K - VP9 2160p/4K}
  \label{fig:VP9Twitch4K}
\end{subfigure}%
\hspace{.01\textwidth}
\begin{subfigure}[T]{.22\textwidth}
  \centering
  \vspace{-3mm}
  \includegraphics[width=0.90\linewidth]{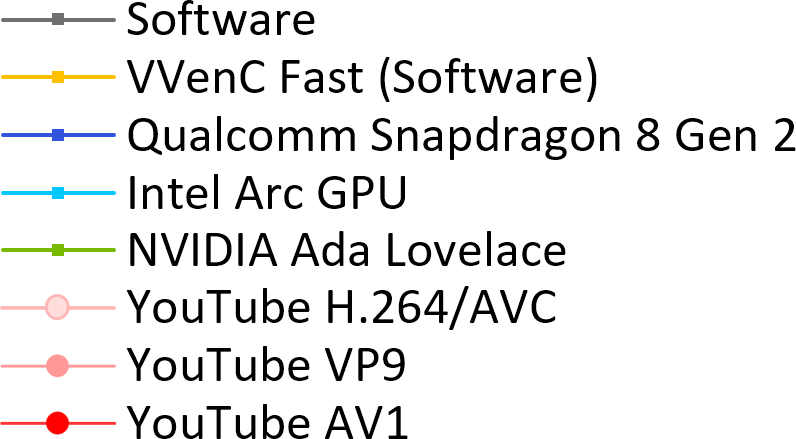}
  \vspace{-1.5mm}
  
\end{subfigure}
\vspace{-2mm}
\caption{Rate-Distortion Curves based on VMAF score of various hardware encoders compared to software encoders and YouTube transcoding when encoding ITE 4K and Twitch dataset at 1080p and 2160p, and ITE 8K dataset at 4320p.}
\label{fig:RDCurves}
\vspace{-6mm}
\end{figure}

At the resolution of 4320p/8K, it was found that the lead gain by hardware encoders started to decline. Intel still maintains a VMAF lead of 0.50, which is a similar RD performance to 1080p. However, NVIDIA now falls behind the software encoder for the first time, trailing behind by 0.29 points. Nevertheless, Qualcomm closed its gap a little bit more, losing by 4.63 points at this resolution. However, when considering PSNR, the results suggested that Qualcomm's disadvantage actually grew larger as it's now losing by 1.45 dB. Intel and NVIDIA's advantages over the software encoders are also gone at this resolution, yielding a negligible difference in PSNR. To better visualize the result, the RD curve based on the VMAF score of every test scenario is also provided in \cref{fig:RDCurves}. Here, it can be clearly seen that Intel encoders have a slight lead over software encoders in most cases. NVIDIA encoder managed to roughly match the RD performance of the software counterpart. However, Qualcomm encoders clearly fall behind by a significant margin.

Finally, when considering the intermediate resolutions including 720p/HD and 1440p/HD, It was found that Intel performed better than NVIDIA from a VMAF score perspective. However, from the PSNR perspective, the advantage shifted to NVIDIA. This suggested that each encoder might be optimized for a different metric. SSIM suggested that both encoders performed similarly in this regard as Intel performed better in 21 test cases compared to 15 won by NVIDIA. The SSIM results suggested that NVIDIA performed better in the older H.264/AVC codec, while Intel performed better in modern codecs like H.265/HEVC and AV1.

\begin{table*}[!tbp]
\setstretch{0.85}
\captionsetup{justification=centering}
\caption{VMAF score comparison between hardware generations relative to the latest generation from each manufacturer. DNE = Didn't Experiment.}
\vspace{-2mm}
\centering
\label{tab:Generation}
\resizebox{18cm}{!}{\begin{tabular}{@{}ll ccc ccc ccc ccc ccc@{}}
\toprule

\multirow{4}{*}{Manufacturer}&\multirow{4}{*}{Hardware}&\multicolumn{3}{c}{ITE 8K}& \multicolumn{6}{c}{ITE 4K} & \multicolumn{6}{c}{Twitch 4K}    \\\cmidrule(lr){3-5}\cmidrule(lr){6-11}\cmidrule(l){12-17}

&&\multicolumn{3}{c}{H.265/HEVC}&\multicolumn{3}{c}{H.264/AVC}&\multicolumn{3}{c}{H.265/HEVC}&\multicolumn{3}{c}{H.264/AVC}&\multicolumn{3}{c}{H.265/HEVC}\\\cmidrule(lr){3-5}\cmidrule(lr){6-8}\cmidrule(lr){9-11}\cmidrule(lr){12-14}\cmidrule(l){15-17}
 &  & Low & Medium & High& Low & Medium & High& Low & Medium & High& Low & Medium & High& Low & Medium & High\\
\midrule
NVIDIA&Ada Lovelace (Latest)&0.000&0.000&0.000&0.000&0.000&0.000&0.000&0.000&0.000&0.000&0.000&0.000&0.000&0.000&0.000               \\
&Ampere&-0.001&0.000&0.001&-0.340&-0.069&0.096&0.066&0.048&0.037&0.073&0.015&-0.020&0.040&0.032&0.027                                \\
&Turing&-0.001&0.000&0.001&-0.343&-0.070&0.097&0.066&0.048&0.037&0.072&0.014&-0.021&0.040&0.032&0.027                                \\
&Pascal&\textbf{\textcolor{red}{-0.324}}&\textbf{\textcolor{red}{-0.399}}&\textbf{\textcolor{red}{-0.445}}&-0.347&-0.692&-0.902&\textbf{\textcolor{red}{-0.410}}&\textbf{\textcolor{red}{-0.442}}&\textbf{\textcolor{red}{-0.462}}&-0.628&-0.691&-0.729&0.316&0.067&\textbf{\textcolor{red}{-0.084}}                       \\
&Maxwell 2nd Generation&N/A&N/A&N/A&-0.347&-0.692&-0.902&-0.410&-0.442&-0.462&-0.609&-0.684&-0.730&0.551&0.346&0.221                 \\
&Maxwell 1st Generation&N/A&N/A&N/A&-0.356&-0.694&-0.900&N/A&N/A&N/A&-0.612&-0.684&-0.728&N/A&N/A&N/A                                \\\midrule
Intel&Arc GPU (Latest)&0.000&0.000&0.000&0.000&0.000&0.000&0.000&0.000&0.000&0.000&0.000&0.000&0.000&0.000&0.000                     \\
&12-13th Generation&-1.511&-0.804&-0.373&-0.943&-0.394&-0.059&-2.018&-1.022&-0.414&-0.211&0.215&0.476&0.026&-0.294&-0.489            \\
&6-9th Generation&N/A&N/A&N/A&-1.278&-0.677&-0.310&\textbf{\textcolor{red}{-5.270}}&\textbf{\textcolor{red}{-4.008}}&\textbf{\textcolor{red}{-3.237}}&-0.469&0.129&0.495&\textbf{\textcolor{red}{-1.891}}&\textbf{\textcolor{red}{-1.979}}&\textbf{\textcolor{red}{-2.033}}                      \\\midrule
Qualcomm&SD 8 Gen 2 (Latest)&0.000&0.000&0.000&0.000&0.000&0.000&0.000&0.000&0.000&0.000&0.000&0.000&0.000&0.000&0.000               \\
&SD 8 Gen 1&0.905&0.666&0.520&0.196&0.210&0.218&-0.189&-0.139&-0.109&DNE&DNE&DNE&DNE&DNE&DNE                                         \\
&SD 888&N/A&N/A&N/A&-1.809&-0.338&0.560&0.039&0.354&0.546&DNE&DNE&DNE&DNE&DNE&DNE                                                    \\

\bottomrule
\end{tabular}}
\vspace{-3mm}
\end{table*}

\begin{table*}[!tbp]
\setstretch{0.85}
\caption{VMAF score achieved by each hardware encoder in three codecs: H.264/AVC, H.265/HEVC, and AV1 compared to Software Encoders at three different bitrate ranges. Note that software encoder experiments were not performed for intermediate resolutions, 720p and 1440p, but the result is provided for reference purposes.}
\vspace{-2mm}
\centering
\label{tab:BitrateComparedVMAF}
\resizebox{18cm}{!}{\begin{tabular}{@{}lccccccccccccccccccc@{}}
\toprule

\multirow{4}{*}{Resolution}&\multirow{4}{*}{Encoder}& \multicolumn{9}{c}{ITE (4K/8K)} & \multicolumn{9}{c}{Twitch (4K)}\\\cmidrule(lr){3-11}\cmidrule(l){12-20}
& & \multicolumn{3}{c}{H.264/AVC} & \multicolumn{3}{c}{H.265/HEVC} & \multicolumn{3}{c}{AV1} & \multicolumn{3}{c}{H.264/AVC} & \multicolumn{3}{c}{H.265/HEVC} & \multicolumn{3}{c}{AV1} \\\cmidrule(lr){3-5}\cmidrule(lr){6-8}\cmidrule(lr){9-11}\cmidrule(lr){12-14}\cmidrule(lr){15-17}\cmidrule(l){18-20}
&&Low&Medium&High&Low&Medium&High&Low&Medium&High&Low&Medium&High&Low&Medium&High&Low&Medium&High\\
\midrule
720p/HD & Intel&\textbf{\textcolor{blue}{67.76}}&\textbf{\textcolor{blue}{77.55}}&\textbf{\textcolor{blue}{83.54}}&\textbf{\textcolor{blue}{73.22}}&\textbf{\textcolor{blue}{81.06}}&\textbf{\textcolor{blue}{85.85}}&\textbf{\textcolor{blue}{74.26}}&\textbf{\textcolor{blue}{81.64}}&\textbf{\textcolor{blue}{86.15}}&\textbf{\textcolor{blue}{74.69}}&\textbf{\textcolor{blue}{85.74}}&\textbf{\textcolor{blue}{92.49}}&\textbf{\textcolor{blue}{78.48}}&86.68&91.68&\textbf{\textcolor{blue}{82.65}}&\textbf{\textcolor{blue}{90.08}}&\textbf{\textcolor{blue}{94.62}}       \\
& NVIDIA&66.41&76.71&83.01&70.60&78.30&83.00&71.34&79.29&84.14&72.40&83.84&90.83&76.62&\textbf{\textcolor{blue}{88.02}}&\textbf{\textcolor{blue}{94.98}}&79.81&87.65&92.44      
\\\midrule
1080p/FHD 
&Software&63.86&74.49&\textbf{\textcolor{blue}{80.99}}&69.50&76.14&80.20&68.49&76.06&80.67&70.68&\textbf{\textcolor{blue}{82.24}}&\textbf{\textcolor{blue}{89.31}}&76.62&83.95&88.42&79.29&86.03&90.15    \\
&Intel&64.42&73.93&79.74&\textbf{\textcolor{blue}{70.16}}&\textbf{\textcolor{blue}{77.75}}&\textbf{\textcolor{blue}{82.38}}&\textbf{\textcolor{blue}{71.05}}&\textbf{\textcolor{blue}{78.12}}&\textbf{\textcolor{blue}{82.44}}&\textbf{\textcolor{blue}{71.07}}&81.90&88.52&72.84&83.84&\textbf{\textcolor{blue}{90.55}}&\textbf{\textcolor{blue}{79.58}}&\textbf{\textcolor{blue}{86.77}}&\textbf{\textcolor{blue}{91.17}}       \\
&NVIDIA&\textbf{\textcolor{blue}{64.72}}&\textbf{\textcolor{blue}{74.70}}&80.79&69.29&76.49&80.89&69.94&77.46&82.05&70.59&81.89&88.79&\textbf{\textcolor{blue}{76.86}}&\textbf{\textcolor{blue}{84.88}}&89.78&78.19&85.92&90.65      \\
&Qualcomm&52.84&66.62&75.04&64.33&72.38&77.29&N/A&N/A&N/A&51.45&72.21&84.88&71.24&80.59&86.29&N/A&N/A&N/A                \\\midrule
1440p/QHD
&Intel&\textbf{\textcolor{blue}{69.84}}&\textbf{\textcolor{blue}{78.53}}&\textbf{\textcolor{blue}{83.83}}&\textbf{\textcolor{blue}{74.89}}&\textbf{\textcolor{blue}{81.86}}&\textbf{\textcolor{blue}{86.12}}&\textbf{\textcolor{blue}{76.09}}&\textbf{\textcolor{blue}{82.35}}&\textbf{\textcolor{blue}{86.18}}&\textbf{\textcolor{blue}{78.44}}&\textbf{\textcolor{blue}{88.38}}&\textbf{\textcolor{blue}{94.44}}&81.85&\textbf{\textcolor{blue}{90.98}}&\textbf{\textcolor{blue}{96.56}}&\textbf{\textcolor{blue}{86.42}}&\textbf{\textcolor{blue}{92.53}}&\textbf{\textcolor{blue}{96.26}}       \\
&NVIDIA&68.58&77.90&83.59&72.58&79.40&83.57&73.35&80.34&84.61&75.96&86.39&92.75&\textbf{\textcolor{blue}{82.08}}&89.16&93.48&83.23&89.96&94.07      \\\midrule
2160p/4K &Software&68.10&78.16&\textbf{\textcolor{blue}{84.31}}&73.86&79.53&82.99&75.71&\textbf{\textcolor{blue}{82.00}}&\textbf{\textcolor{blue}{85.84}}&77.08&87.07&93.17&83.44&89.41&93.05&\textbf{\textcolor{blue}{86.38}}&\textbf{\textcolor{blue}{91.80}}&\textbf{\textcolor{blue}{95.11}}    \\
&Intel&70.45&78.20&82.93&\textbf{\textcolor{blue}{75.74}}&\textbf{\textcolor{blue}{81.60}}&\textbf{\textcolor{blue}{85.18}}&\textbf{\textcolor{blue}{75.90}}&81.62&85.11&\textbf{\textcolor{blue}{79.21}}&\textbf{\textcolor{blue}{88.18}}&93.65&83.38&\textbf{\textcolor{blue}{90.88}}&\textbf{\textcolor{blue}{95.46}}&86.26&91.73&95.08       \\
&NVIDIA&\textbf{\textcolor{blue}{71.10}}&\textbf{\textcolor{blue}{79.19}}&84.13&73.70&79.90&83.68&74.95&81.03&84.74&78.85&88.14&\textbf{\textcolor{blue}{93.81}}&\textbf{\textcolor{blue}{83.83}}&90.34&94.32&85.16&91.30&95.05      \\
&Qualcomm&62.15&72.34&78.57&69.03&75.94&80.16&N/A&N/A&N/A&61.62&79.24&89.99&78.90&86.89&91.77&N/A&N/A&N/A                \\\midrule
4320p/8K &Software&N/A&N/A&N/A&81.76&86.55&89.48&\textbf{\textcolor{blue}{83.60}}&\textbf{\textcolor{blue}{88.25}}&\textbf{\textcolor{blue}{91.09}}&N/A&N/A&N/A&N/A&N/A&N/A&N/A&N/A&N/A  \\
&Intel&N/A&N/A&N/A&\textbf{\textcolor{blue}{82.75}}&\textbf{\textcolor{blue}{87.87}}&\textbf{\textcolor{blue}{91.00}}&83.51&87.95&90.66&N/A&N/A&N/A&N/A&N/A&N/A&N/A&N/A&N/A     \\
&NVIDIA&N/A&N/A&N/A&81.39&86.75&90.02&82.40&87.62&90.81&N/A&N/A&N/A&N/A&N/A&N/A&N/A&N/A&N/A    \\
&Qualcomm&N/A&N/A&N/A&75.93&82.10&85.86&N/A&N/A&N/A&N/A&N/A&N/A&N/A&N/A&N/A&N/A&N/A&N/A        \\

\bottomrule
\end{tabular}}
\vspace{-3mm}
\end{table*}

\begin{table*}[!tbp]
\setstretch{0.85}
\caption{PSNR (dB) achieved by each hardware encoder in three codecs: H.264/AVC, H.265/HEVC, and AV1 compared to Software Encoders at three different bitrate ranges.}
\vspace{-2mm}
\centering
\label{tab:BitrateComparedPSNR}
\resizebox{18cm}{!}{\begin{tabular}{@{}lccccccccccccccccccc@{}}
\toprule

\multirow{4}{*}{Resolution}&\multirow{4}{*}{Encoder}& \multicolumn{9}{c}{ITE (4K/8K)} & \multicolumn{9}{c}{Twitch (4K)}\\\cmidrule(lr){3-11}\cmidrule(l){12-20}
& & \multicolumn{3}{c}{H.264/AVC} & \multicolumn{3}{c}{H.265/HEVC} & \multicolumn{3}{c}{AV1} & \multicolumn{3}{c}{H.264/AVC} & \multicolumn{3}{c}{H.265/HEVC} & \multicolumn{3}{c}{AV1} \\\cmidrule(lr){3-5}\cmidrule(lr){6-8}\cmidrule(lr){9-11}\cmidrule(lr){12-14}\cmidrule(lr){15-17}\cmidrule(l){18-20}
&&Low&Medium&High&Low&Medium&High&Low&Medium&High&Low&Medium&High&Low&Medium&High&Low&Medium&High\\
\midrule
720p/HD&Intel&\textbf{\textcolor{blue}{33.85}}&35.97&37.27&34.69&36.70&37.92&\textbf{\textcolor{blue}{35.06}}&36.90&38.03&35.24&37.39&38.69&35.02&37.62&39.21&36.50&38.33&39.45        \\
&NVIDIA&\textbf{\textcolor{blue}{33.85}}&\textbf{\textcolor{blue}{36.30}}&\textbf{\textcolor{blue}{37.80}}&\textbf{\textcolor{blue}{34.90}}&\textbf{\textcolor{blue}{36.96}}&\textbf{\textcolor{blue}{38.23}}&34.99&\textbf{\textcolor{blue}{37.06}}&\textbf{\textcolor{blue}{38.32}}&\textbf{\textcolor{blue}{35.52}}&\textbf{\textcolor{blue}{38.15}}&\textbf{\textcolor{blue}{39.76}}&\textbf{\textcolor{blue}{36.63}}&\textbf{\textcolor{blue}{38.87}}&\textbf{\textcolor{blue}{40.24}}&\textbf{\textcolor{blue}{36.76}}&\textbf{\textcolor{blue}{38.97}}&\textbf{\textcolor{blue}{40.32}}              \\\midrule
1080p/FHD&Software&32.82&35.11&36.50&\textbf{\textcolor{blue}{34.84}}&36.27&37.14&34.78&36.39&37.37&34.29&36.63&38.07&35.75&37.40&38.40&\textbf{\textcolor{blue}{36.56}}&\textbf{\textcolor{blue}{38.22}}&\textbf{\textcolor{blue}{39.23}}   \\
&Intel&33.70&35.68&36.89&34.67&\textbf{\textcolor{blue}{36.46}}&\textbf{\textcolor{blue}{37.55}}&\textbf{\textcolor{blue}{34.94}}&\textbf{\textcolor{blue}{36.58}}&\textbf{\textcolor{blue}{37.58}}&\textbf{\textcolor{blue}{34.91}}&37.07&38.39&34.52&37.00&38.51&36.19&38.01&39.12               \\
&NVIDIA&\textbf{\textcolor{blue}{33.73}}&\textbf{\textcolor{blue}{35.85}}&\textbf{\textcolor{blue}{37.14}}&34.80&36.44&37.44&34.87&36.53&37.54&34.89&\textbf{\textcolor{blue}{37.18}}&\textbf{\textcolor{blue}{38.58}}&\textbf{\textcolor{blue}{35.92}}&\textbf{\textcolor{blue}{37.81}}&\textbf{\textcolor{blue}{38.96}}&36.06&37.91&39.05              \\
&Qualcomm&31.32&33.88&35.45&33.35&35.09&36.15&N/A&N/A&N/A&31.58&35.00&37.08&34.51&36.49&37.71&N/A&N/A&N/A                        \\\midrule
1440p/QHD&Intel&35.95&37.51&38.46&35.17&36.86&37.90&\textbf{\textcolor{blue}{36.35}}&37.71&38.54&37.57&39.80&41.17&37.60&40.19&41.77&39.07&40.86&41.96      \\
&NVIDIA&\textbf{\textcolor{blue}{36.18}}&\textbf{\textcolor{blue}{37.68}}&\textbf{\textcolor{blue}{38.60}}&\textbf{\textcolor{blue}{35.21}}&\textbf{\textcolor{blue}{37.14}}&\textbf{\textcolor{blue}{38.32}}&36.28&\textbf{\textcolor{blue}{37.78}}&\textbf{\textcolor{blue}{38.69}}&\textbf{\textcolor{blue}{37.72}}&\textbf{\textcolor{blue}{40.35}}&\textbf{\textcolor{blue}{41.96}}&\textbf{\textcolor{blue}{38.95}}&\textbf{\textcolor{blue}{41.06}}&\textbf{\textcolor{blue}{42.35}}&\textbf{\textcolor{blue}{39.17}}&\textbf{\textcolor{blue}{41.24}}&\textbf{\textcolor{blue}{42.51}}              \\\midrule
2160p/4K&Software&33.68&35.25&36.21&35.32&36.22&36.77&35.56&\textbf{\textcolor{blue}{36.59}}&\textbf{\textcolor{blue}{37.22}}&38.19&40.59&42.05&40.03&41.57&42.51&\textbf{\textcolor{blue}{40.81}}&\textbf{\textcolor{blue}{42.45}}&43.45    \\
&Intel&34.68&35.90&36.64&\textbf{\textcolor{blue}{35.43}}&\textbf{\textcolor{blue}{36.44}}&\textbf{\textcolor{blue}{37.06}}&\textbf{\textcolor{blue}{35.57}}&36.54&37.14&39.12&41.38&42.76&39.38&41.87&\textbf{\textcolor{blue}{43.39}}&40.54&42.36&43.47               \\
&NVIDIA&\textbf{\textcolor{blue}{34.81}}&\textbf{\textcolor{blue}{36.08}}&\textbf{\textcolor{blue}{36.86}}&35.29&36.32&36.95&35.47&36.48&37.09&\textbf{\textcolor{blue}{39.15}}&\textbf{\textcolor{blue}{41.55}}&\textbf{\textcolor{blue}{43.01}}&\textbf{\textcolor{blue}{40.11}}&\textbf{\textcolor{blue}{42.06}}&43.25&40.46&42.36&\textbf{\textcolor{blue}{43.53}}              \\
&Qualcomm&33.39&34.93&35.87&34.37&35.48&36.15&N/A&N/A&N/A&35.79&39.32&41.47&38.82&40.85&42.10&N/A&N/A&N/A                        \\\midrule
4320p/8K&Software&N/A&N/A&N/A&\textbf{\textcolor{blue}{36.85}}&38.10&38.86&\textbf{\textcolor{blue}{37.16}}&\textbf{\textcolor{blue}{38.50}}&\textbf{\textcolor{blue}{39.32}}&N/A&N/A&N/A&N/A&N/A&N/A&N/A&N/A&N/A                            \\
&Intel&N/A&N/A&N/A&36.72&\textbf{\textcolor{blue}{38.19}}&\textbf{\textcolor{blue}{39.08}}&37.12&38.38&39.15&N/A&N/A&N/A&N/A&N/A&N/A&N/A&N/A&N/A                                       \\
&NVIDIA&N/A&N/A&N/A&36.69&38.14&39.02&36.87&38.27&39.13&N/A&N/A&N/A&N/A&N/A&N/A&N/A&N/A&N/A                                      \\
&Qualcomm&N/A&N/A&N/A&35.15&36.68&37.62&N/A&N/A&N/A&N/A&N/A&N/A&N/A&N/A&N/A&N/A&N/A&N/A                                          \\

\bottomrule
\end{tabular}}
\vspace{-3mm}
\end{table*}

\begin{table*}[!tbp]
\setstretch{0.85}
\caption{SSIM achieved by each hardware encoder in three codecs: H.264/AVC, H.265/HEVC, and AV1 compared to Software Encoders at three different bitrate ranges.}
\vspace{-2mm}
\centering
\label{tab:BitrateComparedSSIM}
\resizebox{18cm}{!}{\begin{tabular}{@{}lccccccccccccccccccc@{}}
\toprule

\multirow{4}{*}{Resolution}&\multirow{4}{*}{Encoder}& \multicolumn{9}{c}{ITE (4K/8K)} & \multicolumn{9}{c}{Twitch (4K)}\\\cmidrule(lr){3-11}\cmidrule(l){12-20}
& & \multicolumn{3}{c}{H.264/AVC} & \multicolumn{3}{c}{H.265/HEVC} & \multicolumn{3}{c}{AV1} & \multicolumn{3}{c}{H.264/AVC} & \multicolumn{3}{c}{H.265/HEVC} & \multicolumn{3}{c}{AV1} \\\cmidrule(lr){3-5}\cmidrule(lr){6-8}\cmidrule(lr){9-11}\cmidrule(lr){12-14}\cmidrule(lr){15-17}\cmidrule(l){18-20}
&&Low&Medium&High&Low&Medium&High&Low&Medium&High&Low&Medium&High&Low&Medium&High&Low&Medium&High\\
\midrule
720p/HD&Intel&0.9255&0.9532&0.9701&\textbf{\textcolor{blue}{0.9423}}&\textbf{\textcolor{blue}{0.9624}}&\textbf{\textcolor{blue}{0.9747}}&\textbf{\textcolor{blue}{0.9468}}&\textbf{\textcolor{blue}{0.9650}}&\textbf{\textcolor{blue}{0.9761}}&0.9446&0.9700&\textbf{\textcolor{blue}{0.9856}}&\textbf{\textcolor{blue}{0.9320}}&0.9628&0.9816&\textbf{\textcolor{blue}{0.9591}}&\textbf{\textcolor{blue}{0.9757}}&\textbf{\textcolor{blue}{0.9859}}       \\
&NVIDIA&\textbf{\textcolor{blue}{0.9304}}&\textbf{\textcolor{blue}{0.9578}}&\textbf{\textcolor{blue}{0.9746}}&0.9408&0.9613&0.9738&0.9426&0.9626&0.9747&\textbf{\textcolor{blue}{0.9537}}&\textbf{\textcolor{blue}{0.9731}}&0.9849&0.9317&\textbf{\textcolor{blue}{0.9635}}&\textbf{\textcolor{blue}{0.9829}}&0.9549&0.9730&0.9841             \\\midrule
1080p/FHD&Software&0.9309&0.9596&0.9772&\textbf{\textcolor{blue}{0.9517}}&0.9677&0.9775&0.9476&0.9645&0.9749&\textbf{\textcolor{blue}{0.9394}}&0.9692&\textbf{\textcolor{blue}{0.9873}}&\textbf{\textcolor{blue}{0.9618}}&0.9766&0.9856&0.9646&0.9778&0.9859  \\
&Intel&0.9336&0.9582&0.9733&0.9513&\textbf{\textcolor{blue}{0.9682}}&\textbf{\textcolor{blue}{0.9785}}&\textbf{\textcolor{blue}{0.9545}}&\textbf{\textcolor{blue}{0.9694}}&\textbf{\textcolor{blue}{0.9785}}&0.9377&0.9660&0.9833&0.9525&0.9736&0.9864&\textbf{\textcolor{blue}{0.9662}}&\textbf{\textcolor{blue}{0.9795}}&\textbf{\textcolor{blue}{0.9876}}              \\
&NVIDIA&\textbf{\textcolor{blue}{0.9381}}&\textbf{\textcolor{blue}{0.9620}}&\textbf{\textcolor{blue}{0.9767}}&0.9487&0.9659&0.9764&0.9498&0.9666&0.9768&0.9373&\textbf{\textcolor{blue}{0.9668}}&0.9848&0.9607&\textbf{\textcolor{blue}{0.9767}}&0.9864&0.9617&0.9765&0.9856             \\
&Qualcomm&0.9162&0.9520&0.9739&0.9410&0.9622&0.9752&N/A&N/A&N/A&0.9026&0.9550&0.9869&0.9589&0.9761&\textbf{\textcolor{blue}{0.9866}}&N/A&N/A&N/A                             \\\midrule
1440p/QHD&Intel&0.9603&0.9761&0.9858&\textbf{\textcolor{blue}{0.9735}}&\textbf{\textcolor{blue}{0.9835}}&\textbf{\textcolor{blue}{0.9896}}&\textbf{\textcolor{blue}{0.9756}}&\textbf{\textcolor{blue}{0.9840}}&\textbf{\textcolor{blue}{0.9891}}&0.9797&0.9899&\textbf{\textcolor{blue}{0.9961}}&0.9669&0.9842&0.9948&\textbf{\textcolor{blue}{0.9854}}&\textbf{\textcolor{blue}{0.9919}}&\textbf{\textcolor{blue}{0.9959}}     \\
&NVIDIA&\textbf{\textcolor{blue}{0.9652}}&\textbf{\textcolor{blue}{0.9800}}&\textbf{\textcolor{blue}{0.9891}}&0.9718&0.9823&0.9888&0.9727&0.9827&0.9888&\textbf{\textcolor{blue}{0.9827}}&\textbf{\textcolor{blue}{0.9909}}&0.9959&\textbf{\textcolor{blue}{0.9677}}&\textbf{\textcolor{blue}{0.9854}}&\textbf{\textcolor{blue}{0.9963}}&0.9838&0.9911&0.9955             \\\midrule
2160p/4K&Software&0.9689&0.9834&\textbf{\textcolor{blue}{0.9922}}&0.9809&0.9883&0.9927&0.9834&0.9892&\textbf{\textcolor{blue}{0.9928}}&0.9790&\textbf{\textcolor{blue}{0.9912}}&\textbf{\textcolor{blue}{0.9986}}&\textbf{\textcolor{blue}{0.9894}}&0.9942&0.9972&0.9906&0.9947&0.9972   \\
&Intel&0.9726&0.9836&0.9903&\textbf{\textcolor{blue}{0.9839}}&\textbf{\textcolor{blue}{0.9894}}&\textbf{\textcolor{blue}{0.9928}}&\textbf{\textcolor{blue}{0.9839}}&\textbf{\textcolor{blue}{0.9894}}&\textbf{\textcolor{blue}{0.9928}}&0.9786&0.9902&0.9974&0.9889&\textbf{\textcolor{blue}{0.9945}}&\textbf{\textcolor{blue}{0.9979}}&\textbf{\textcolor{blue}{0.9915}}&\textbf{\textcolor{blue}{0.9953}}&\textbf{\textcolor{blue}{0.9977}}              \\
&NVIDIA&\textbf{\textcolor{blue}{0.9755}}&\textbf{\textcolor{blue}{0.9859}}&\textbf{\textcolor{blue}{0.9922}}&0.9798&0.9874&0.9921&0.9813&0.9877&0.9916&\textbf{\textcolor{blue}{0.9794}}&0.9910&0.9980&0.9892&0.9944&0.9976&0.9902&0.9947&0.9975             \\
&Qualcomm&0.9668&0.9823&0.9917&0.9776&0.9868&0.9924&N/A&N/A&N/A&0.9578&0.9842&0.9960&0.9887&0.9942&0.9976&N/A&N/A&N/A                             \\\midrule
4320p/8K&Software&N/A&N/A&N/A&0.9925&0.9958&0.9978&0.9946&0.9966&\textbf{\textcolor{blue}{0.9978}}&N/A&N/A&N/A&N/A&N/A&N/A&N/A&N/A&N/A                                       \\
&Intel&N/A&N/A&N/A&\textbf{\textcolor{blue}{0.9935}}&\textbf{\textcolor{blue}{0.9962}}&\textbf{\textcolor{blue}{0.9979}}&\textbf{\textcolor{blue}{0.9949}}&\textbf{\textcolor{blue}{0.9967}}&0.9977&N/A&N/A&N/A&N/A&N/A&N/A&N/A&N/A&N/A                                                  \\
&NVIDIA&N/A&N/A&N/A&0.9923&0.9955&0.9975&0.9934&0.9957&0.9971&N/A&N/A&N/A&N/A&N/A&N/A&N/A&N/A&N/A                                                 \\
&Qualcomm&N/A&N/A&N/A&0.9916&0.9951&0.9973&N/A&N/A&N/A&N/A&N/A&N/A&N/A&N/A&N/A&N/A&N/A&N/A                                                        \\

\bottomrule
\end{tabular}}
\vspace{-3mm}
\end{table*}

\begin{table*}[!tbp]
\setstretch{0.85}
\caption{Relative VMAF score compared to Software Encoders}
\vspace{-2mm}
\centering
\label{tab:BitrateRelativeVMAF}
\resizebox{18cm}{!}{\begin{tabular}{@{}lccccccccccccccccccc@{}}
\toprule

\multirow{4}{*}{Resolution}&\multirow{4}{*}{Encoder}& \multicolumn{9}{c}{ITE (4K/8K)} & \multicolumn{9}{c}{Twitch (4K)}\\\cmidrule(lr){3-11}\cmidrule(l){12-20}
& & \multicolumn{3}{c}{H.264/AVC} & \multicolumn{3}{c}{H.265/HEVC} & \multicolumn{3}{c}{AV1} & \multicolumn{3}{c}{H.264/AVC} & \multicolumn{3}{c}{H.265/HEVC} & \multicolumn{3}{c}{AV1} \\\cmidrule(lr){3-5}\cmidrule(lr){6-8}\cmidrule(lr){9-11}\cmidrule(lr){12-14}\cmidrule(lr){15-17}\cmidrule(l){18-20}
&&Low&Medium&High&Low&Medium&High&Low&Medium&High&Low&Medium&High&Low&Medium&High&Low&Medium&High\\
\midrule
1080p/FHD &Intel&0.56&-0.56&-1.24&0.66&1.61&2.19&2.56&2.07&1.77&0.39&-0.34&-0.79&-3.78&-0.11&2.13&0.29&0.74&1.02        \\
&NVIDIA&0.87&0.20&-0.20&-0.21&0.35&0.70&1.45&1.41&1.38&-0.09&-0.36&-0.52&0.24&0.93&1.36&-1.10&-0.11&0.50      \\
&Qualcomm&\textbf{\textcolor{red}{-11.02}}&\textbf{\textcolor{red}{-7.88}}&\textbf{\textcolor{red}{-5.95}}&\textbf{\textcolor{red}{-5.17}}&-3.77&-2.91&N/A&N/A&N/A&\textbf{\textcolor{red}{-19.23}}&\textbf{\textcolor{red}{-10.04}}&-4.43&\textbf{\textcolor{red}{-5.38}}&-3.36&-2.12&N/A&N/A&N/A  \\
\midrule
2160p/4K & Intel&1.87&2.07&2.19&2.35&0.03&-1.38&0.19&-0.38&-0.73&2.13&1.11&0.49&-0.06&1.47&2.41&-0.13&-0.07&-0.03       \\
&NVIDIA&-0.16&0.37&0.69&3.00&1.03&-0.17&-0.76&-0.97&-1.10&1.76&1.07&0.65&0.39&0.93&1.26&-1.23&-0.50&-0.06     \\
&Qualcomm&-4.84&-3.59&-2.83&\textbf{\textcolor{red}{-5.95}}&\textbf{\textcolor{red}{-5.82}}&\textbf{\textcolor{red}{-5.74}}&N/A&N/A&N/A&\textbf{\textcolor{red}{-15.46}}&\textbf{\textcolor{red}{-7.83}}&-3.18&-4.53&-2.52&-1.28&N/A&N/A&N/A    \\
\midrule
4320p/8K & Intel&N/A&N/A&N/A&0.99&1.32&1.52&-0.09&-0.30&-0.43&N/A&N/A&N/A&N/A&N/A&N/A&N/A&N/A&N/A                       \\
&NVIDIA&N/A&N/A&N/A&-0.37&0.20&0.55&-1.20&-0.63&-0.28&N/A&N/A&N/A&N/A&N/A&N/A&N/A&N/A&N/A                     \\
&Qualcomm&N/A&N/A&N/A&\textbf{\textcolor{red}{-5.83}}&-4.46&-3.62&N/A&N/A&N/A&N/A&N/A&N/A&N/A&N/A&N/A&N/A&N/A&N/A                       \\
\bottomrule
\end{tabular}}
\vspace{-4mm}
\end{table*}

Despite the newer hardware generation promising a better RD performance, it was found that the performance stays relatively the same among the past few generations of hardware as seen in Table \ref{tab:Generation}. As the result suggested, the differences usually occur when a new encoder's features are being implemented such as B-Frame encoding support for H.265/HEVC codec in NVIDIA Turing GPU, which yields a slight bit-saving compared to Pascal GPU that lacks the support. Another case is the introduction of Intel Arc dedicated GPU, which features a completely re-designed GPU core, resulting in major changes in the hardware encoder, resulting in a major performance uplift over the previous generations, yielding over 5 points improvement in VMAF score compared to older GPUs.

\begin{figure}[t!]
\centering
\begin{subfigure}{.24\textwidth}
  \centering\includesvg[width=0.95\linewidth,inkscapelatex=false]{ITE4K_Codecs.svg}
  \vspace{-1.5mm}
  \caption{ITE 4K}
  \label{fig:ITE4K_Codecs}
\end{subfigure}%
\begin{subfigure}{.24\textwidth}
  \centering\includesvg[width=0.95\linewidth,inkscapelatex=false]{Twitch4K_Codecs.svg}
  \vspace{-1.5mm}
  \caption{Twitch 4K}
  \label{fig:Twitch4K_Codecs}
\end{subfigure}\\
\begin{subfigure}{.24\textwidth}
  \centering\includesvg[width=0.95\linewidth,inkscapelatex=false]{ITE1080p_Codecs.svg}
  \vspace{-1.5mm}
  \caption{ITE 4K (1080p)}
  \label{fig:ITE1080p_Codecs}
\end{subfigure}%
\begin{subfigure}{.24\textwidth}
  \centering\includesvg[width=0.95\linewidth,inkscapelatex=false]{Twitch1080p_Codecs.svg}
  \vspace{-1.5mm}
  \caption{Twitch 1080p}
  \label{fig:Twitch1080p_Codecs}
\end{subfigure}\\
\begin{subfigure}[T]{.24\textwidth}
  \centering
  \vspace{-3mm}
\centering\includesvg[width=0.95\linewidth,inkscapelatex=false]{ITE8K_Codecs.svg}
  \vspace{-1.5mm}
  \caption{ITE 8K}
  \label{fig:ITE8K_Codecs}
\end{subfigure}%
\hspace{.01\textwidth}
\begin{subfigure}[T]{.22\textwidth}
  \centering
  \vspace{-3mm}
  \includegraphics[width=0.60\linewidth]{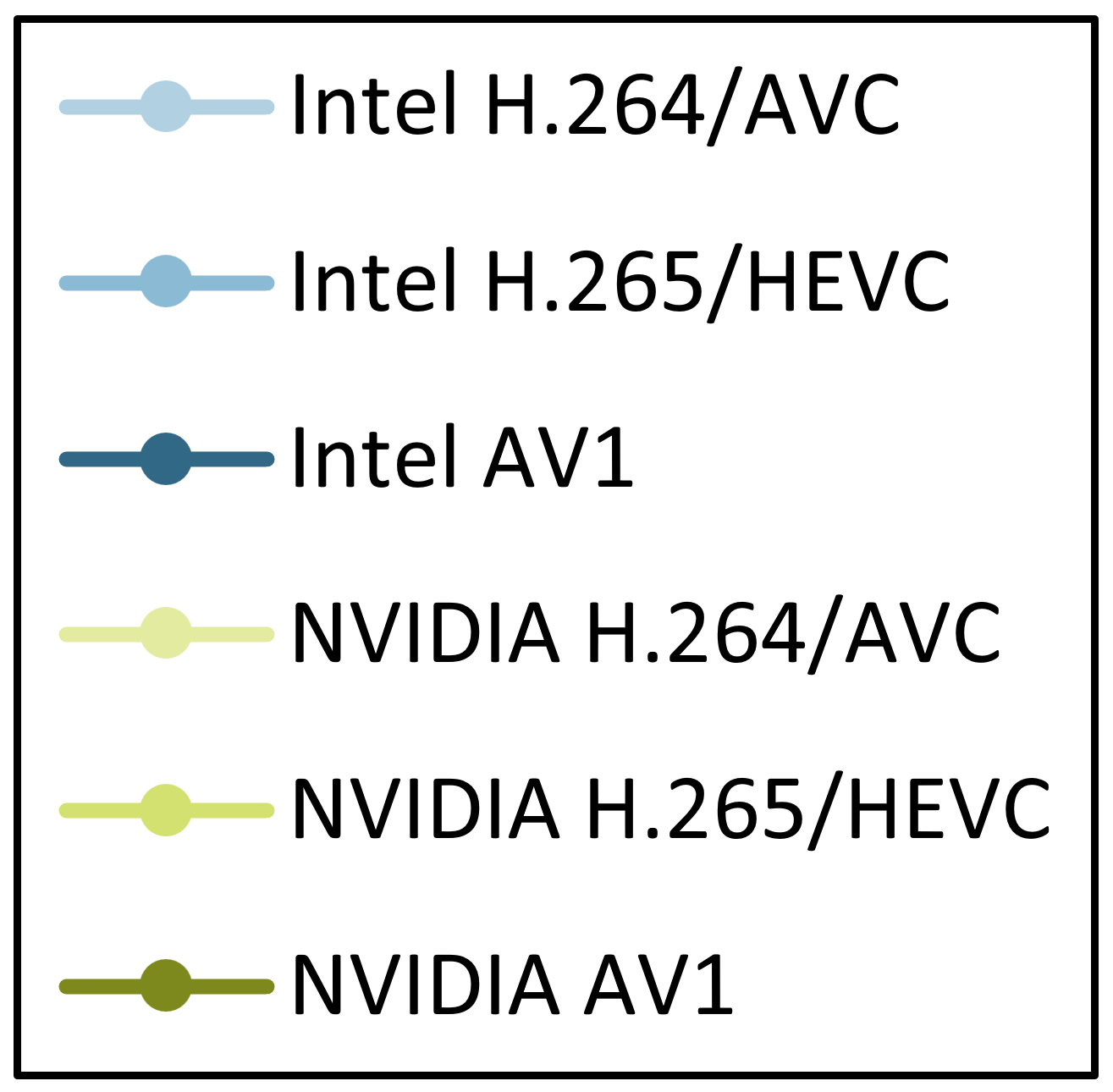}
  \vspace{-1.5mm}
  
\end{subfigure}
\vspace{-2mm}
\caption{Rate-Distortion Curves of three different codecs: H.264/AVC, H.265/HEVC, and AV1 as encoded by Intel and NVIDIA encoders when performing encoding on various datasets at various resolutions.}
\label{fig:RDCurvesCodecs}
\vspace{-2mm}
\end{figure}

Another case that resulted in a significantly improved RD performance is when a new codec is being implemented. \cref{fig:RDCurvesCodecs} shows the RD curves of three codecs as supported by the latest generation Intel and NVIDIA GPU. By opting for newer codecs with additional bit-saving features like larger macro block size in H.265 and film grain synthesis in AV1, the RD performance can be significantly improved over the older codecs. From the figure, it can be seen that by utilizing H.265/HEVC for video encoding tasks, a major RD performance advantage can be gained compared to using H.264/AVC. It should be noted that using newer codecs may impact the compatibility as older devices may not support the newer codecs, however, since most online streaming platforms usually perform transcoding, this shouldn't be an issue in most cases. Finally, it should be noted that the RD performance improvement over H.265/HEVC brought by AV1 codecs by this generation of hardware is relatively small compared to the difference between H.265/HEVC and H.264/AVC. Results from \cref{tab:EncodingSpeed4K} suggested that the main focus of the manufacturers seems to be on increasing the encoding throughput over maximizing the efficiency as this is more important to realize real-time encoding at higher resolutions like 4320p/8K.

\subsection{Bitrate for YouTube Live-streaming}

Lastly, the bitrate required by each encoder to match the quality of YouTube transcoding was found. Typically, only the popular channel will get VP9 and AV1 encoding, while smaller channel will only get their video transcode to H.264/AVC. However, if the input video resolution is 1440p or higher, YouTube will provide VP9 encoding regardless of the popularity. By downloading the uploaded YouTube video and calculating the VMAF score, it was found that YouTube AV1 and VP9 yield a very similar quality, while YouTube H.264/AVC quality is way worse. By using the VMAF score calculated from YouTube's transcode of Twitch dataset, and the RD curve obtained from the previous subsection, which can also be seen in Figure \ref{fig:RDCurves}, the bitrate required to match YouTube transcoding quality at each resolution when the encoder from NVIDIA and Intel is configured with live-streaming parameters can be found. The Twitch dataset was chosen because most of the user who uses the hardware encoder on GPU are usually game streamers and VTubers intend to stream their video games. The result in \cref{tab:YouTubeCompared} can be used by streamers and VTubers as a guideline when configuring their streaming software. It was found that to match the quality of YouTube H.264/AVC requires about 50\% less bitrate at 2160p/4K compared to their VP9 and AV1 encoding, but the difference was reduced to around 42\% at 1080p/FHD. The results also suggested that a 10-20\% higher bitrate is required for the NVIDIA encoder to match the quality of the Intel encoder given the same resolution and codec. \looseness=-1

\begin{table}[!tbp]
\setstretch{0.85}
\caption{Bitrate required to match YouTube Transcoding Quality at each resolution and codec (kbps)}
\vspace{-1.5mm}
\centering
\label{tab:YouTubeCompared}
\resizebox{8.7cm}{!}{\begin{tabular}{@{}lccccccccc@{}}
\toprule

\multirow{2.5}{*}{Resolution}&\multirow{2.5}{*}{Codec}&\multirow{2.5}{*}{VMAF}& \multicolumn{3}{c}{Intel} & \multicolumn{3}{c}{NVIDIA}    \\\cmidrule(lr){4-6}\cmidrule(l){7-9}
& &  & AVC & HEVC & AV1 & AVC & HEVC & AV1 \\
\midrule
YT 720p/HD&AVC&75.82&1568&1390&768&1798&1164&1023\\
&VP9&85.13&2821&2457&1842&3171&2570&2342\\
&AV1&84.96&2791&2432&1813&3138&2533&2307\\\midrule
YT 1080p/FHD&AVC&80.42&5329&4720&3166&5355&3979&3569\\
&VP9&87.95&8652&7608&6571&8521&7656&7037\\
&AV1&88.03&8697&7647&6622&8563&7709&7088\\\midrule
YT 1440p/QHD&VP9&88.53&11854&9728&7432&13533&11018&10111\\
&AV1&86.81&10507&8532&6109&12063&9302&8462\\\midrule
YT 2160p/4K&VP9&93.03&42767&35814&34588&42304&39092&35664\\
&AV1&93.92&45799&38870&38692&45225&43000&39454\\

\bottomrule
\end{tabular}}
\vspace{-6mm}
\end{table}

\section{Conclusions}

In this paper, the performance of hardware encoders on modern GPUs was evaluated in many aspects, including RD performance, encoding throughput, and power consumption. Additionally, the RD performance was compared across multiple hardware generations as well as different codecs on the same hardware using two datasets ITE 4K/8K and Twitch via three objective quality assessment metrics including PSNR, SSIM, and VMAF. The RD curve was also found for various experiment case scenarios. Comparisons between the dataset and real-world content were also made to provide the readers with helpful context of our evaluation and can help them easily apply the results in their applications. Finally, the bitrate required for each hardware encoder, when used in live encoding settings, to match the quality of YouTube transcoding at every resolution, and codecs were found to help with the decision-making process of game streamers and VTubers when configuring their live encoder parameters. By presenting the data in an easy-to-understand format, the results in this research can also be used by the end-users outside this field of study, adding value to this work.

It was found that hardware encoders yield a similar encoding throughput in all codecs when compared to software encoders with very fast or preset 9 running on high-end desktop CPUs while consuming significantly less power. Remarkably, the results show that Intel QuickSync yields better RD performance when compared to the best real-time-encoding-applicable preset of each codec that can encode 1080p60 video in real-time on a high-end laptop CPU, saving about 5\% of data across the board, while NVIDIA NVENC performed is slightly behind, requiring about 10\% more bits to reach the same quality as Intel. However, Qualcomm Snapdragon SoCs need 50\% higher bitrate to match the software counterpart, falling significantly behind the software counterpart. Additionally, it was found that there is a negligible difference in RD performances between hardware released during the past six years, instead, the RD performance improvement is gained from the use of new codecs implemented in newer hardware generations. Therefore, with capable hardware streaming to a supported online streaming platform, these newer video codecs should be used for real-time streaming as it can yield a major RD performance uplift over the older hardware and codecs.

Lastly, it was found that YouTube H.264/AVC encoding yields significantly worse quality than VP9 and AV1 counterparts, requiring about 46\% lower bitrate from hardware encoders to match its quality. Since YouTube supports H.264/AVC, H.265/HEVC, and AV1 codec for live encoding, streamers should use the newest codec supported by their hardware to maximize the quality while saving their network bandwidth. While end-users may choose bitrate just to match YouTube H.264/AVC encoding quality to save the bandwidth, popular streamers should aim for VP9 or AV1 quality as their stream will be transcoded to VP9 and AV1 codec after the live is completed. This will guarantee that the audience will receive the highest video quality possible, maximizing the Quality of Experience (QoE).

\section{Future Work}

While the encoding speed of hardware encoders at the slowest preset is sufficient for 2160p/4K resolution and lower, which is currently commonly used by most of the streamers, it's not adequate to handle the 4320p/8K resolution, especially at high frame rate (60-120 fps). In the future, various different hardware encoders preset will be tested and evaluated as this is one of the keys towards real-time 8K content creation and distribution. Additionally, the latest features designed to improve the encoding throughput such as recently introduced NVIDIA Split-Frame Encoding \cite{monteiro_yaghoubian_2024} as part of their Ada Lovelace GPUs will also be evaluated to better understand the RD performance and encoding speed impacts. Finally, subjective evaluations may be incorporated to offer insights into user-perceived Quality of Experience (QoE) as it is known that objective metrics might not be a perfect representation of the experience perceived by the actual audience.

\section*{Acknowledgement}




This work was partially supported by Hoso Bunka Foundation and Kayamori Foundation of Information Science Advancement. Additionally, the authors gratefully thank the \textbf{\#VTuberTH} community for the inspiration of this research, especially \textbf{Maylyn} and \textbf{MIINA}, who drove this research using the power of \textit{V siang siang}.




%
\vspace{-1mm}
\setstretch{0.95}
\Urlmuskip=0mu plus 1mu\relax

\bibliographystyle{IEEEtran}
\bibliography{b_reference}

\end{document}